\documentclass[aps,prd,showpacs,amssymb,nofootinbib]{revtex4}
\usepackage{graphicx}
\usepackage[usenames,dvipsnames]{color}

\begin{document}
\title{Asymptotics of Regulated Field Commutators for Einstein-Rosen Waves}

\author{J. Fernando \surname{Barbero G.}}
\email[]{fbarbero@iem.cfmac.csic.es} \affiliation{Instituto de
Estructura de la Materia, Centro de F\'{\i}sica Miguel A.
Catal\'{a}n, C.S.I.C., Serrano 121, 28006 Madrid, Spain}
\author{Guillermo A. \surname{Mena Marug\'an}}
\email[]{mena@iem.cfmac.csic.es} \affiliation{Instituto de
Estructura de la Materia, Centro de F\'{\i}sica Miguel A. Catal\'{a}n,
C.S.I.C., Serrano 121, 28006 Madrid, Spain}
\author{Eduardo J. \surname{S. Villase\~nor}}
\email[]{ejsanche@math.uc3m.es} \affiliation{Departamento de
Matem\'aticas, Escuela Polit\'ecnica Superior, Universidad Carlos
III de Madrid, Avda. de la Universidad 30, 28911 Legan\'es, Spain}

\date{December 3, 2004}

\begin{abstract}
We discuss the asymptotic behavior of regulated field commutators
for linearly polarized, cylindrically symmetric gravitational
waves and the mathematical techniques needed for this analysis. We
concentrate our attention on the effects brought about by the
introduction of a physical cut-off in the study of the
microcausality of the model and describe how the different
physically relevant regimes are affected by its presence.
Specifically we discuss how genuine quantum gravity effects can be
disentangled from those originating in the introduction of a
regulator.

\end{abstract}


\maketitle

\section{\label{Intro}Introduction}

Linearly polarized cylindrical waves, also known as Einstein-Rosen
waves \cite{Einstein-Rosen,Kuchar:1971xm}, provide a symmetry
reduction of general relativity that can be used as a test bed for
the quantization of the theory. This system displays several
interesting features that contribute to its relevance. On one hand
it has an infinite number of local degrees of freedom and, hence,
it is a genuine quantum field theory (in contradistinction to
other symmetry reductions, such as Bianchi models, that have a
finite number of global degrees of freedom). On the other, the
system is tractable both classically and quantum mechanically,
thus allowing to derive exact consequences independent of any
approximation scheme
\cite{Ashtekar:1996bb,Ashtekar:1996yk,Mad:2000,Angulo:2000ad,
BarberoG.:2003ye,BarberoG.:2004}. The main reason behind this
success and tractability is the fact that the gravitational
degrees of freedom of the model are encoded in an free, massless,
axially symmetric, scalar field that evolves in an auxiliary
Minkowskian background.

In previous papers we have analyzed the issue of microcausality in
this system{;} in particular{,} we have studied in detail the
smearing of {light} cones owing to the quantization of the
gravitational field \cite{BarberoG.:2003ye,BarberoG.:2004}. The
main tool for this type of analysis is the study {in vacuo} of the
field commutator evaluated at different spacetime points. As is
well known the commutator of quantum fields reflects the causal
structure of spacetime (Minkowskian spacetime in ordinary
perturbative quantum field theory) in the sense that the quantum
fields in spatially separated spacetime points commute. This is
true for {all standard types} of quantum {fields,} i.e. scalar,
fermion, or vector fields --though issues related to gauge
invariance must be carefully considered in this last case--. In
the specific model that we are interested in{,} gauge invariance
has been discussed in {Ref. \cite{Kouletsis:2003hj}}. The authors
of {that} paper conclude that it is correct to use the
Ashtekar-Pierri gauge fixed action, written in terms of the
axially symmetric scalar field, to derive gauge invariant
information about the model{.}

In a {recent work} \cite{BarberoG.:2004} we {discussed the}
situation {when} no cut-off is introduced {in the system, studying
the unregulated commutator}. The main results {reached were} the
following{.} First, one can clearly see that light cones are
smeared by quantum gravity effects; in fact it is possible to
obtain a quantitative measure of this smearing and {show} how
sharp light cones are recovered in the limit of large distances as
compared to the natural length scale of the model --the Planck
length--. It is also interesting to point out that the asymptotic
behavior of the commutator in the different physically relevant
regimes strongly depends on the causal relationship between the
different spacetime points involved. Second{, one finds} a
singularity structure in the {commutator} that differs from {that
of} the free theory; in particular the field commutator for equal
values of the radial coordinate $R$ is singular. Finally one
observes that{,} in the case when one of the spacetime points that
{appear} in the commutator corresponds to the symmetry axis{,}
there are quantum effects that persist for large values of the
difference of the time {coordinates}. This effect is reminiscent
of the large quantum gravity effects {first} discussed by Ashtekar
{\cite{Ashtekar:1996yk,Angulo:2000ad,Gambini:1997,Tiglio:1999}}.

The purpose of this paper is to study how the conclusions of Ref.
\cite{BarberoG.:2004} are changed by the introduction of a
cut-off. As is well {known} regulators are {generally} necessary
in order to have well defined quantum field theories. One can
justify its use, for example, by noticing that the action of the
field operator on the vacuum in a Fock space is not a vector in
the Hilbert space because it has infinite norm. In order to have a
well defined action of the field operator one regulates it by
introducing smearing functions that render the norms of these
states finite. The problem then {consists in} removing these
regulators (or rather showing that the physical results are
independent of them).

In principle{,} it is possible to argue that the results derived
in the absence of regulators somehow approximate those derived
after their introduction; this is straightforward to see in the
case of cut-offs. In the presence of a cut-off $\Lambda${,} the
improper integrals that define the field commutator become proper
because the integration region is a closed interval $[0,\Lambda]$.
For a given value of the parameters that appear in the integral
(involving the values of the spacetime coordinates of the quantum
fields and the gravitational constant) it is always possible to
choose a value for $\Lambda$ such that the integral with the
cut-off is well approximated by the integral extended to
$[0,\infty)$.

Of course it is conceivable that the cut-off is not just a
mathematical device but rather a physical scale defining a
fundamental limit for the resolution of our measurements. If
spacetime becomes discrete at short distances (such as the Planck
length) the continuum spacetime picture breaks {down} and,
certainly, it would be difficult to justify the extension of the
integrals involved in the definition of field commutators (or $S$
matrix elements, for that matter) to infinite intervals in momenta
(inverse length). Our point of view here is that the introduction
of a cut-off can mimic some of the effects appearing after a
successful quantization of gravity (for example, in the loop
quantum gravity approach) and hence we plan to study its effect
within the {consistent} framework provided by the Einstein-Rosen
waves. It is also interesting to point out here that the cut-off
by itself can produce some of the effects expected from quantum
gravity. In particular, it is possible to show that light cones
are also smeared by cut-offs\footnote{This is easy to understand
for ordinary field theories in Minkowski spacetime because
cut-offs break Lorentz invariance in these models.}. In our
opinion this makes it necessary to study in detail how the effects
of the cut-off and quantum gravity can be disentangled.

The paper is organized as follows. After this introduction{,} we
briefly review the main results about microcausality in quantum
cylindrical gravitational waves and introduce the commutators that
we will discuss in the rest of the paper. {We} will {then} study
the field commutators in the presence of a cut-off with the help
of the asymptotic techniques already employed in Ref.
\cite{BarberoG.:2004}. Here the situation is simpler because we
will only have to consider integrals over closed intervals. We
will discuss one by one the asymptotic behaviors in the different
parameters involved. In {Sec. VI} we will derive {a} power series
expansion in the gravitational constant for the commutator in the
presence of a cut-off in the spirit of ordinary perturbative
quantum field theory{,} and discuss the uniform convergence of
this series under appropriate conditions on the cut-off in {Sec.}
VII. We end the paper with a discussion of our results and our
conclusions.

\section{\label{commutator}The Field Commutator}

Einstein-Rosen waves describe topologically trivial spacetimes
with two linearly independent, commuting, spacelike, and
hypersurface orthogonal Killing vector fields \cite{Romano:1996ep,
Einstein-Rosen, Kuchar:1971xm} endowed with a metric that can be
written as
\begin{equation}
ds^2=e^{\gamma-\psi}(-dT^2+dR^2) +e^{-\psi}R^2d\theta^2+e^\psi
dZ^2.\label{metric}
\end{equation}
Here we use the coordinates
$(T,R,\theta,Z),\,\,T\in\mathbb{R},\,\,R\in[0,\infty),\,\,
\theta\in[0,2\pi),\,\,Z\in\mathbb{R}$, and $\psi$ and $\gamma$ are
functions only of $R$ and $T$. The Einstein field equations are
very simple. The scalar field $\psi$  satisfies the wave equation
for a massless, axially symmetric scalar field in three dimensions
$$
\partial_T^2\psi-\partial_R^2\psi-\frac{1}{R}\partial_R\psi=0
$$
and the function $\gamma$ can be expressed in terms of this field
\cite{Ashtekar:1996bb, Angulo:2000ad} as
$$
\gamma(R)=\frac{1}{2}\int^R_0 d\bar{R} \,\bar{R}\, \left[
(\partial_T\psi)^2+(\partial_{\bar{R}}\psi)^2 \right].
$$
We will use in the following a system of units such that
$c=\hbar=1$ and define $G\equiv\hbar G_3$, where $G_3$ denotes the
gravitational constant per unit length in the direction of the
symmetry axis\footnote{Notice that $G$ has dimensions of length in
this system of units.}. The function $\gamma(R)$ (apart from a
factor of $8G$) has a simple physical interpretation:  it is the
energy of the scalar field in a ball of radius $R$ whereas
$\gamma_{\infty}$ denotes $\lim_{R\rightarrow\infty}\gamma(R)$
(the energy of the whole two-dimensional flat space). It is also
possible to show \cite{Ashtekar:1996bb, BarberoG.:2003ye} that
$\gamma_\infty/8G$ coincides with the Hamiltonian $H_0$ of the
system obtained by a linearization of the metric (\ref{metric}).

In order to have a unit asymptotic timelike Killing vector and a
physical notion of energy (per unit length) we introduce the
coordinates $(t,R,\theta,Z)$ defined by
$T=e^{-\gamma_{\infty}/2}t$. In these coordinates the metric takes
the form \cite{Kuchar:1971xm, Romano:1996ep}
$$
ds^2=e^{\gamma-\psi}(-e^{-\gamma_\infty}dt^2+dR^2)
+e^{-\psi}R^2d\theta^2+e^\psi dZ^2.
$$
By taking a sufficiently fast fall-off for $\psi$ as $R\rightarrow
\infty$, this metric describes asymptotically flat {cylindrical}
spacetimes such that $\partial_t$ is a unit timelike Killing
vector in the asymptotic region. In the 2+1 dimensional framework
these spacetimes are asymptotically flat at spacelike and null
infinities \cite{Ashtekar:1997cm, Ashtekar:1996cd} (the
appropriate introduction of null infinity will be needed in order
to study the $S$ matrix of the model). It is also worthwhile
noting that these spacetimes have a non-zero deficit angle.

The Einstein field equations can be obtained from a Hamiltonian
action principle \cite{Ashtekar:1994ds, Varadarajan:1995hw,
Romano:1996ep}. {A remarkable} (and useful) feature of the
physical Hamiltonian $H$ (associated with the physical time $t$)
is the fact that it is a function of the {Hamiltonian}
corresponding to the free scalar field, $H_0$:
\begin{eqnarray*}
H=E(H_0)=\frac{1}{4G}(1-e^{-4GH_0}).
\end{eqnarray*}

In terms of the $T$-time and imposing regularity at the axis $R=0$
\cite{Ashtekar:1996bb}, the classical solutions for the field
$\psi$ can be written as
\begin{eqnarray*}
\psi(R,T)=\sqrt{4G}\int_0^\infty \!\!dk\, J_0(Rk)
\left[A(k)e^{-ikT} +A^\dagger(k)e^{ikT}\right]
\end{eqnarray*}
where $A(k)$ and its complex conjugate $A^\dagger(k)$ are fixed by
the initial conditions. The free Hamiltonian $H_0$ can be written
now as
\begin{eqnarray*}
{\frac{\gamma_{\infty}}{8G}}=H_0=\int_0^{\infty}dk\,kA^{\dagger}(k)A(k).
\end{eqnarray*}
Using this expression, we obtain the $t$-evolution of the field
\begin{eqnarray*}
\psi_E(R,t)= \sqrt{4G}\int_0^\infty\!\! dk\,
J_0(Rk)\left[A(k)e^{-ikte^{-\gamma_\infty/2}}
\!+\!A^\dagger(k)e^{ikte^{-\gamma_\infty/2}}\right].\;
\end{eqnarray*}

The quantization can be carried out in the usual way by
introducing a Fock space where $\hat{\psi}(R,0)$, the quantum
counterpart of $\psi(R,0)$, is the operator-valued distribution
\cite{Reed:1975uy} given by\footnote{In the main body of the paper
we will be employing regulated fields; we will refer to the
unregulated ones only in this section{,} so we label them with a
$\Lambda=\infty$ tag.}
\begin{eqnarray}
\hat{\psi}(R,0;\Lambda=\infty)\!=
\!\hat{\psi}_E(R,0;\Lambda=\infty)\!=\!\sqrt{4G}\int_0^\infty
\!\!\!dk\,
J_0(Rk)\!\left[\hat{A}(k)\!+\!\hat{A}^\dagger(k)\right]\!.
\label{psi0}
\end{eqnarray}
Its action on Fock {space} is determined by {those} of
$\hat{A}(k)$ and $\hat{A}^\dagger(k)$ --the annihilation and
creation operators-- with non-vanishing commutators given by $
[\hat{A}(k_1),\hat{A}^\dagger(k_2)]=\delta(k_1,k_2)$.

We can regulate the field by introducing suitable functions $g(k)$
that render finite the norms of the states obtained by acting with
the quantum field on Fock space vectors. In the following we will
make the simplest choice $g(k)=\chi_{[0,\Lambda_k]}(k)$ (here
$\chi_{[a,b]}$ denotes the characteristic function of the interval
$[a,b]$). By doing this the integration region in (\ref{psi0})
{becomes} compact and we have
\begin{eqnarray}
\hat{\psi}(R,0)\!=\!\hat{\psi}_E(R,0)\!=\!\sqrt{4G}\int_0^{\Lambda_k}
\!\!\!dk\,
J_0(Rk)\!\left[\hat{A}(k)\!+\!\hat{A}^\dagger(k)\right]\!.
\label{psi0lambda}
\end{eqnarray}

Evolution in $T$ is given by the unitary operator
$\hat{U}_0(T)=\exp(-iT\hat{H}_0)$ where
\begin{eqnarray*}\label{qham0}
\hat{H}_0=\int^{\infty}_0dk\,k\,\hat{A}^{\dagger}(k)\hat{A}(k)
\end{eqnarray*}
is the quantum Hamiltonian operator of a {three-dimensional},
axially symmetric scalar field. The cut-off regulated quantum
scalar field in the Heisenberg picture is hence given by
\begin{eqnarray}
\hspace*{-.3cm}\hat{\psi}(R,T)=
\hat{U}^\dagger_0(T)\hat{\psi}(R,0)\hat{U}_0(T)=\sqrt{4G}\int_0^{\Lambda_k}\!\!
dk\,
J_0(Rk)\!\left[\hat{A}(k)e^{-ikT}\!+\!\hat{A}^\dagger(k)e^{ikT}\right]
\!.\hspace*{.3cm}\nonumber
\end{eqnarray}
If we describe the evolution in our model in terms of the physical
time $t$, the quantum Hamiltonian is
$\hat{H}=E(\hat{H}_0)=\frac{1}{4G}(1-e^{-4G\hat{H}_0})$ and
unitary evolution is given by $\hat{U}(t)=\exp(-it\hat{H})$. With
this time evolution the annihilation and creation operators in the
Heisenberg picture are
\begin{eqnarray}
\hat{A}_E(k,t)\!\!&\equiv&\!\!\hat{U}^\dagger(t)\hat{A}(k)\hat{U}(t)=
\exp\!\left[-itE(k)e^{-4G\hat{H}_0}\right]\!\hat{A}(k),\nonumber\\
\hat{A}_E^\dagger(k,t)\!\!&=&\!\hat{A}^\dagger(k)\,\exp\!\left[itE(k)
e^{-4G\hat{H}_0}\right],\nonumber
\end{eqnarray}
where $E(k)=\frac{1}{4G}(1-e^{-4Gk})$, and the regulated field
operator evolved with the physical Hamiltonian [that we denote as
$\hat{\psi}_E(R,t)$] is given by
\begin{eqnarray*}
\hat{\psi}_E(R,t)=\sqrt{4G}\int_0^{\Lambda_k}\!\!\!dk
\,J_0(Rk)\left[\hat{A}_E(k,t)+\hat{A}_E^\dagger(k,t)\right].
\end{eqnarray*}
The field commutator $\big[\hat{\psi}_E(R_1,t_1),
\hat{\psi}_E(R_2,t_2)\big]$ can be computed from these expressions
\cite{BarberoG.:2003ye}. Since we are dealing with an effectively
interacting theory this operator is not proportional to the
identity in the Fock space basis that we are using and, hence, we
have to consider its matrix elements. As in previous work we will
concentrate on the vacuum expectation value. If a cut-off
$\Lambda_k$ is introduced{,} this is given by
\begin{eqnarray}
\frac{1}{8iG}\langle 0|\,
[\hat{\psi}_E(R_1,t_1),\hat{\psi}_E(R_2,t_2)]\,|0\rangle=
\int_0^{\Lambda_k}\!\!\!\!\!\!dk\,J_0(R_1 k)J_0(R_2
k)\sin\left[\frac{t_2-t_1}{4G}(1\!-\!e^{-4Gk})\right]\!{,}
\hspace*{.4cm}&&\label{integral}
\end{eqnarray}
that can be seen to depend on the time coordinates {only} through
their difference $t_2-t_1$ --which we will assume in the following
to be positive--. Notice that it depends symmetrically on $R_1$
and $R_2$. The functional dependence in $G$ is less trivial, a
fact that requires especial attention when studying the limit in
which the relevant lengths and time differences are much larger
than the Planck length \cite{BarberoG.:2004}.

{It is} convenient to refer the dimensional parameters of these
integrals to another length scale{,} that we choose as $R_1$. We
hence introduce $R_2=\rho R_1$, $t_2-t_1={R_1}\tau$, and
$\lambda=R_1/(4G)$ and rewrite (\ref{integral}) as
\begin{equation}
\frac{1}{8iG}\langle 0|\,
[\hat{\psi}_E(R_1,t_1),\hat{\psi}_E(R_2,t_2)]\,|0\rangle=
\frac{\lambda}{R_1}\Im\bigg\{\int_0^{\Lambda_q}dqJ_0(\lambda
q)J_0(\rho \lambda
q)e^{i\tau\lambda(1-e^{-q})}\bigg\}\label{newintegral}
\end{equation}
after introducing the new variable $k=q/(4G)$. Here $\Im$ denotes
the imaginary part and $\Lambda_q=4G\Lambda_k$.

Physically{,} the cut-off $\Lambda_k$ ({which in principle has the
dimensionality of an} inverse length) could be interpreted as
having its origin in the existence of a minimum length. This comes
out naturally in loop quantum gravity where space is discrete and
the {area} and volume operators {have minimum eigenvalues of the
order of} the Planck area and volume{,} respectively. In fact{,}
the existence of a {minimum} length (of the size of the Planck
{scale}) {can be considered} a generic feature of essentially
every quantum {theory} \cite{Garay:1995en}. The interpretation of
the adimensional cut-off $\Lambda_q$ {would follow} from that of
$\Lambda_k$, so it {may be} reasonable to expect that it {be} a
number of order unity; nevertheless we will treat it as a free
parameter in the following.

\section{\label{ro}Asymptotic behavior in $\rho$}

Let us start by considering the behavior of (\ref{newintegral})
when the parameter $\rho$ grows to infinite or approaches
$\rho=0$. This integral can be written as a standard $h$-transform
\cite{Handels} by the change of variables $t=q\lambda$. The most
convenient way to get its asymptotic behavior in
$\rho\rightarrow\infty$ is by rewriting it in the form
\begin{equation}
\frac{1}{R_1}\Im\left[\int_0^{\infty}\!\!\!\!\!dt J_0(\rho
t)J_0(t)e^{i\tau\lambda(1-e^{-t/\lambda})}-\int_{R_1\Lambda_k}^{\infty}
\!\!\!\!\!dt J_0(\rho
t)J_0(t)e^{i\tau\lambda(1-e^{-t/\lambda})}\right]{.}\label{split}
\end{equation}
{One can then} use the asymptotic behavior obtained for the first
integral in Ref. \cite{BarberoG.:2004}, and {find} the asymptotics
of the second {integral} by standard integration by parts
{[employing} the fact that\footnote{The appearance of $1/k$ terms
forces us to split the original integral in two as in
(\ref{split}) to avoid singularities at $k=0$.}
$J_0(k)=-\frac{1}{k}J_0^{\prime}(k)-J_0^{\prime\prime}(k)${]}. By
doing this one gets the following two contributions
$$
\frac{1}{R_1}\Big[\frac{\tau}{2\lambda
\rho^3}+\frac{1}{\rho^5}\Big(\frac{9\tau}{8\lambda}-\frac{3\tau}{8\lambda^3}+
\frac{9\tau^2}{2\lambda}\Big){\Big]}+O(\frac{1}{\rho^7}){,}
$$
$$
\frac{1}{\rho R_1}\Im\Big[J_1(\Lambda_k R_1 \rho)J_0(\Lambda_k
R_1)e^{i\tau\lambda(1-e^{-\Lambda_k
R_1/\lambda})}\Big]+O(\frac{1}{\rho^{5/2}}).
$$

The first one is cut-off independent but subdominant with respect
to the second, hence {we} see that the presence of a cut-off
changes the asymptotic behavior in $\rho$. This is the kind of
behavior that one would expect even in a Lorentz covariant theory
after the introduction of a cut-off because of the breaking of the
Lorentz symmetry. The novel feature here is the presence of
cut-off independent terms. Although the cut-off dependent one
dominates in the asymptotic limit, there may be a transient regime
--whose onset will be controlled by the value of $\Lambda_k$-- in
which the asymptotic behavior is given by the first term. This
will be most evident when $\Lambda_k\rightarrow\infty$.

In the $\rho\rightarrow 0$ limit we get
\begin{equation}
\frac{1}{R_1}\Im\int_0^{R_1\Lambda_k}\!\!\!\!\!dt
J_0(t)e^{i\tau\lambda(1-e^{-t/\lambda})}+O(\rho)=
\frac{\lambda}{R_1}\Im\int_0^{\Lambda_q}\!\!\!\!\!dq J_0(\lambda
q)e^{i\tau\lambda(1-e^{-q})}+O(\rho), \label{r-int-0-lamb}
\end{equation}
as a result of the continuity at $\rho=0$ of the integral defining
the commutator (\ref{newintegral}).

\section{\label{tau}Asymptotic behavior in $\tau$}

The integral in (\ref{newintegral}) has the convenient form of
{an} $h$-transform and, hence, it can be studied by standard
Mellin transform methods \cite{Handels} if the asymptotic
parameter is chosen to be $\rho${;} however this is no longer true
if the asymptotic parameter is taken to be $\tau$ (which
corresponds to considering large separations in the time
coordinates). This fact introduces some mathematical difficulties
in the asymptotic analysis. In this case one has to consider the
cases $\rho=0$ and $\rho\neq0$ separately.

If $\rho=0$ one finds that the asymptotic behavior when
$\tau\rightarrow\infty$ without the cut-off is given by
\cite{BarberoG.:2004}
\begin{equation}
\frac{1}{R_1}\sqrt{\frac{\lambda}{2\pi\log\tau}}
       \Im\Big\{e^{i[\frac{\pi}{4}+\tau\lambda-\lambda\log(\tau\lambda)]}
       e^{\frac{\pi}{2}\lambda}\Gamma(i\lambda)+
       e^{-i[\frac{\pi}{4}-\tau\lambda-\lambda\log(\tau\lambda)]}
       e^{-\frac{\pi}{2}\lambda}\Gamma(-i\lambda)\Big\}+
       O(\frac{1}{\log\tau})\label{tauasimpr0}
\end{equation}
whereas for $\rho\neq0$ we get
\begin{eqnarray}
\frac{1}{2 \pi R_1 \sqrt{\rho}
       \log\tau}
       \Im\Big\{e^{i[\frac{\pi}{2}+\tau\lambda-
       \lambda(1+\rho)\log(\tau\lambda)]}
       e^{\frac{\pi}{2}\lambda(1+\rho)}
       \Gamma[i\lambda(1+\rho)]+e^{-i[\frac{\pi}{2}-
       \tau\lambda-\lambda(1+\rho)\log(\tau\lambda)]}
       e^{-\frac{\pi}{2}\lambda(1+\rho)}
       \Gamma[-i\lambda(1+\rho)]\hspace{6mm}&&\label{tauasimprneq0}\\
       {+}e^{i[\tau\lambda-\lambda(1-\rho)\log(\tau\lambda)]}
       e^{\frac{\pi}{2}\lambda(1-\rho)}
       \Gamma[i\lambda(1-\rho)]+e^{i[\tau\lambda-
       \lambda(\rho-1)\log(\tau\lambda)]}
       e^{\frac{\pi}{2}\lambda(\rho-1)}
       \Gamma[i\lambda(\rho-1)]\Big\}+O
       (\frac{1}{(\log\tau)^2}){.}\nonumber
\end{eqnarray}
The most interesting feature of these expressions is their unusual
dependence on the asymptotic parameter $\tau$; in fact the
dependence on inverse powers of logarithms (especially on the
inverse square root of $\log\tau$) cannot be obtained by direct
application of the usual asymptotic expressions derived by Mellin
transform techniques \cite{BarberoG.:2004}. It is also remarkable
how slowly the commutator decays in $\tau$ --in particular in the
axis $\rho=0$--{,} a fact that is suggestive of the large quantum
gravity effects discussed by {Ashtekar} \cite{Ashtekar:1996yk}.
Outside the axis the decay is {faster} but still quite slow. A
consequence of the different asymptotic behaviors in $\tau$ for
$\rho=0$ and $\rho\neq0$ is the impossibility to recover
(\ref{tauasimpr0}) as the limit when $\rho\rightarrow0$ of
(\ref{tauasimprneq0}). As we can see{,} the frequency of the
oscillations of the commutator in $\tau$ is controlled by
$\lambda$ (proportional to the inverse of $G$) in such a way that
although the amplitude of the oscillations decays very slowly they
will average to zero on scales larger than the Planck length.

When we introduce a cut-off $\Lambda_k${, the above} asymptotic
expressions change to
\begin{equation}
\frac{1}{R_1}\Im\Big\{\frac{i}{\tau}\Big[1-e^{i\tau\lambda(1-e^{-\Lambda_k
R_1/\lambda})}e^{\Lambda_k R_1/\lambda}J_0(\Lambda_k
R_1)J_0(\rho\Lambda_k
R_1)\Big]\Big\}+O(\frac{1}{\tau^2}),\label{tauasimptlamb}
\end{equation}
valid both for $\rho=0$ and $\rho\neq0$. This can be obtained by
straightforward integration by parts. An interesting situation
develops at this point because the asymptotic behavior of the
integral in $\tau$ behaves in a discontinuous way in the cut-off.
In the analysis carried out to study the asymptotic behavior in
$\rho$ we found out that the cut-off dependent term, in spite of
being dominant, goes to zero in the limit
$\Lambda_k\rightarrow\infty$. Here the situation is different{:}
taking now $\Lambda_k\rightarrow\infty$ in (\ref{tauasimptlamb})
does not lead to the asymptotic expressions corresponding to
$\Lambda_k=\infty${. That} is{,} the asymptotic behavior of the
improper integral in (\ref{integral}) is not the limit when
$\Lambda_k\rightarrow\infty$ {of} (\ref{newintegral}). As in the
case of the asymptotics in $\rho$, one expects that there must be
a transient regime in which the behavior in $\tau$ of
(\ref{newintegral}) is given by (\ref{tauasimpr0}) and
(\ref{tauasimprneq0}){. We will not consider here a precise
characterization of this transient behavior for arbitrary values
of the relevant parameters because its main properties can be
conveniently discussed}{, at least for large $\lambda$,} {by
looking at the $\lambda\rightarrow\infty$ asymptotics of the
commutator in the ($\rho,\tau$) plane.}

The $\tau\rightarrow0$ limit is easy to analyze. In fact what we
find, both with and without the cut-off, is that the series
obtained by expanding $e^{i\tau\lambda(1-e^{-t/\lambda})}$ in
powers of $e^{-t/\lambda}$, exchanging integration and infinite
sum, and computing the resulting integrals gives a series that
converges to the value of the commutator.

\section{\label{lambda}Asymptotic behavior in $\lambda$}

The asymptotic behavior in $\lambda$ is studied by following the
procedure described in {Ref.} \cite{BarberoG.:2004}. {It is worth
remarking that the limit $\lambda\rightarrow\infty$ of the
regulated field commutator cannot be identified with that in which
the gravitational constant $G$ vanishes if one admits that the
dimensionful cut-off $\Lambda_k<\infty$ is kept constant {in
principle}. On the contrary, the two limits could be considered
equivalent only under the assumption that $\Lambda_k$ increases as
the inverse of $G$ for small gravitational constant, so that its
dimensionless counterpart $\Lambda_q=4G\Lambda_k$ may remain
fixed.}

{The analysis of the asymptotics in $\lambda$} when the cut-off is
present is simultaneously simpler {in some respects} and more
complicated {in others compared with} the case {when} no cut-off
is introduced. It is simpler because the lengthy analysis needed
to discuss the asymptotics of the improper integral is not
necessary now. It is more complicated in the sense that the final
asymptotic expressions contain additional terms and also because
the number of regions with different $\lambda$-asymptotic regimes
in the $(\rho,\tau)$ plane increases.

We have to consider now the cases $\rho=0$ and $\rho\neq0$
separately. Let us consider first $\rho=0$ and write the r.h.s. of
(\ref{newintegral}) as {\cite{BarberoG.:2004}}
\begin{equation}
\Im\bigg\{-\frac{i\lambda e^{i\tau\lambda}}{2\pi
R_1}\int_0^{\Lambda_q}dq\oint_{\gamma} dt\frac{1}{t}
e^{\lambda\big[\frac{\scriptstyle
q}{\scriptstyle2}(t-\frac{\scriptstyle1}{\scriptstyle t})-i\tau
e^{-q}\big]}{\bigg\}}\label{bintegral}
\end{equation}
after using the usual integral representation for the Bessel
functions $J_n$ ($n=0,\,1,\,\ldots${)}
\begin{eqnarray*}
J_n(z)=\frac{1}{2\pi i}\oint_{\gamma}
\frac{dt}{t^{n+1}}e^{\frac{z}{2}(t-\frac{\scriptstyle
1}{\scriptstyle t})},
\end{eqnarray*}
where $\gamma$ is a closed, positively oriented, simple path in
the complex plane surrounding the origin. Notice that we are
integrating an integrable function in a compact region{,} so we
can write the integrals in any order we want. The asymptotic
analysis of (\ref{bintegral}) can be carried out by following the
same steps as in Ref. \cite{BarberoG.:2004}. As we did there{,} it
is useful to introduce neutralizers to split the integral in three
pieces $I_j$, $j=1,\,2,\,3${,} and choose appropriate contours for
each of them. These integrals are
\begin{eqnarray*}
&&I_j\equiv\Im\Big\{-\frac{i\lambda e^{i\tau\lambda}}{2\pi
R_1}\int_0^{\Lambda_q}dq\oint_{\gamma} dt\,\nu_j(q)\frac{1}{t}
e^{\lambda\big[\frac{\scriptstyle
q}{\scriptstyle2}(t-\frac{\scriptstyle1}{\scriptstyle t})-i\tau
e^{-q}\big]}\Big\}
\end{eqnarray*}
where we have introduced the neutralizer functions $\nu_j(q)$,
$j=1,\,2,\,3${,} satisfying $\nu_1+\nu_2+\nu_3=1$ in
$[0,\Lambda_q]$ and
\begin{eqnarray*}
&&\nu_1(q)=1\quad \textrm{if}\quad q\in[0,\alpha_1],\\
&&\nu_1(q)=0\quad \textrm{if}\quad q\in[\alpha_2,\Lambda_q],\\
&&\nu_2(q)=0\quad \textrm{if}\quad q\in[0,\alpha_1]\cup[\beta_2,\Lambda_q],\\
&&\nu_2(q)=1\quad \textrm{if}\quad q\in[\alpha_2,\beta_1],\\
&&\nu_3(q)=0\quad \textrm{if}\quad q\in[0,\beta_1],\\
&&\nu_3(q)=1\quad \textrm{if}\quad q\in[\beta_2,\Lambda_q],
\end{eqnarray*}
with $0<\alpha_1<\alpha_2<\beta_1<\beta_2<\Lambda_q$ (these
parameters are chosen as in Ref. \cite{BarberoG.:2004}). By doing
this the effective integration regions in $q$ are $[0,\alpha_2]$,
$[\alpha_1,\beta_2]${,} and $[\beta_1,\Lambda_q]$ and the boundary
$q=0$ appears only in the first.

The asymptotics in $\lambda$ of the integral $I_1$ is best
obtained by choosing an integration contour satisfying
$\Re(t-1/t)\leq0$ (that passes necessarily through $t=i$ and
$t=-i$). By using the same method of {Ref.} \cite{BarberoG.:2004}
we see that the first two relevant terms are given by the contour
integrals
$$\frac{1}{\pi R_1}\Im\bigg\{i\oint_{\gamma}\frac{dt}{t^2+2i\tau
t-1} \bigg\}{,}$$
$$
-\frac{1}{2\pi
R_1}\Im\bigg\{\frac{i}{\lambda}\oint_{\gamma}dt\frac{8i\tau
t^2}{(t^2+2i\tau t-1)^3}\bigg\}{,}
$$
whose sum gives
$$
\frac{1}{R_1\sqrt{\tau^2-1}}\quad {\rm for}\,\, \tau>1
$$
and
$$\frac{\tau(1+2\tau^2)}{2R_1\lambda(1-\tau^2)^{5/2}}\quad {\rm for}\,\,
\tau<1.$$ Although the second term will be subdominant with
respect to some of the contributions coming from $I_2$ and
$I_3${,} it improves the approximation of the full commutator
obtained from the asymptotics in $\lambda$ in the region $\tau<1$.

The contribution of $I_3$ to the asymptotics in $\lambda$ is
obtained from the contour integral (corresponding to the boundary
at $q=\Lambda_q$)
$$
\Im\bigg\{\frac{ie^{i\tau\lambda}}{2\pi R_1}
\oint_{\gamma}dt\frac{2}{t} \frac{(t-\frac{1}{\displaystyle
t}+2i\tau
e^{-\Lambda_q})e^{\frac{\lambda{\Lambda_q}}{2}(t-\frac{1}{t})-
i\tau e^{-\Lambda_q}}}{\Lambda_q^2(t+\frac{1}{\displaystyle
t})^2-(t-\frac{1}{\displaystyle t}+2i\tau
e^{-\Lambda_q})^2}\bigg\}.
$$
The asymptotics in $\lambda$ of this integral can be easily
studied by using the method of steepest descents. This gives
$$
\Im\Big\{\frac{2i
e^{i\tau\lambda(1-e^{-\Lambda_q})}}{R_1(\tau^2e^{-2\Lambda_q}-1)
\sqrt{2\pi\lambda\Lambda_q}}\big[i\tau
e^{-\Lambda_q}\sin(\lambda\Lambda_q-\frac{\pi}{4})-
\cos(\lambda\Lambda_q-\frac{\pi}{4})\big]\Big\}.
$$
Finally the integral $I_2$ (for which we choose for $\gamma$ the
curve $|t|=1$) only contributes when the stationary points of the
exponent are in the integration region. This happens only when
$1<\tau<e^{\Lambda_q}$. The contribution to the first relevant
order in $\lambda$ is \cite{BarberoG.:2004}
$$
\Im\bigg\{\frac{1}{R_1}
\frac{e^{i\lambda(\tau-\log\tau-1)}}{\sqrt{\log\tau}}\bigg\}.
$$
Adding up the different terms we get
\begin{eqnarray}
&&\theta(1-\tau)\frac{\tau(1+2\tau^2)}{2R_1\lambda(1-\tau^2)^{5/2}}+
\theta(\tau-1)\frac{1}{R_1\sqrt{\tau^2-1}}+
\theta(\tau-1)\theta(e^{\Lambda_q}-\tau)\Im\bigg\{\frac{1}{R_1}
\frac{e^{i\lambda(\tau-\log\tau-1)}}{\sqrt{\log\tau}}\bigg\}\nonumber\\
&&{+}\Im\Big\{\frac{2i
e^{i\tau\lambda(1-e^{-\Lambda_q})}}{R_1(\tau^2e^{-2\Lambda_q}-1)
\sqrt{2\pi\lambda\Lambda_q}}\big[i\tau
e^{-\Lambda_q}\sin(\lambda\Lambda_q-\frac{\pi}{4})
-\cos(\lambda\Lambda_q-\frac{\pi}{4})\big]\Big\}{,}\label{asinteje}
\end{eqnarray}
where $\theta$ denotes the step function.

{We see that} the final result consists of {several}
contributions: the free commutator for an infinite cut-off, a
$1/\lambda$ correction for $\tau<1$, the term with the
$1/\sqrt{\log\tau}$ dependence for $1<\tau<e^{\Lambda_q}${,} and a
cut-off dependent contribution for all values of $\tau$ that falls
off to zero when $\Lambda_q\rightarrow\infty$. If the cut-off goes
to {infinity} the commutator can be approximated by the one
obtained in Ref. \cite{BarberoG.:2004}{;} however, if it is of
order one (as would be the case if it is defined by the Planck
length){,} then that approximation is no longer valid. Notice that
the values of $\tau\in(1,e^{\Lambda_q})$ are {those} for which the
asymptotics provided by the unregulated commutator is a {correct}
approximation. This is {roughly} the transient region in the
$\tau$ parameter mentioned in the previous subsection.
\begin{figure}
\hspace{0cm}\includegraphics[width=16.5cm]{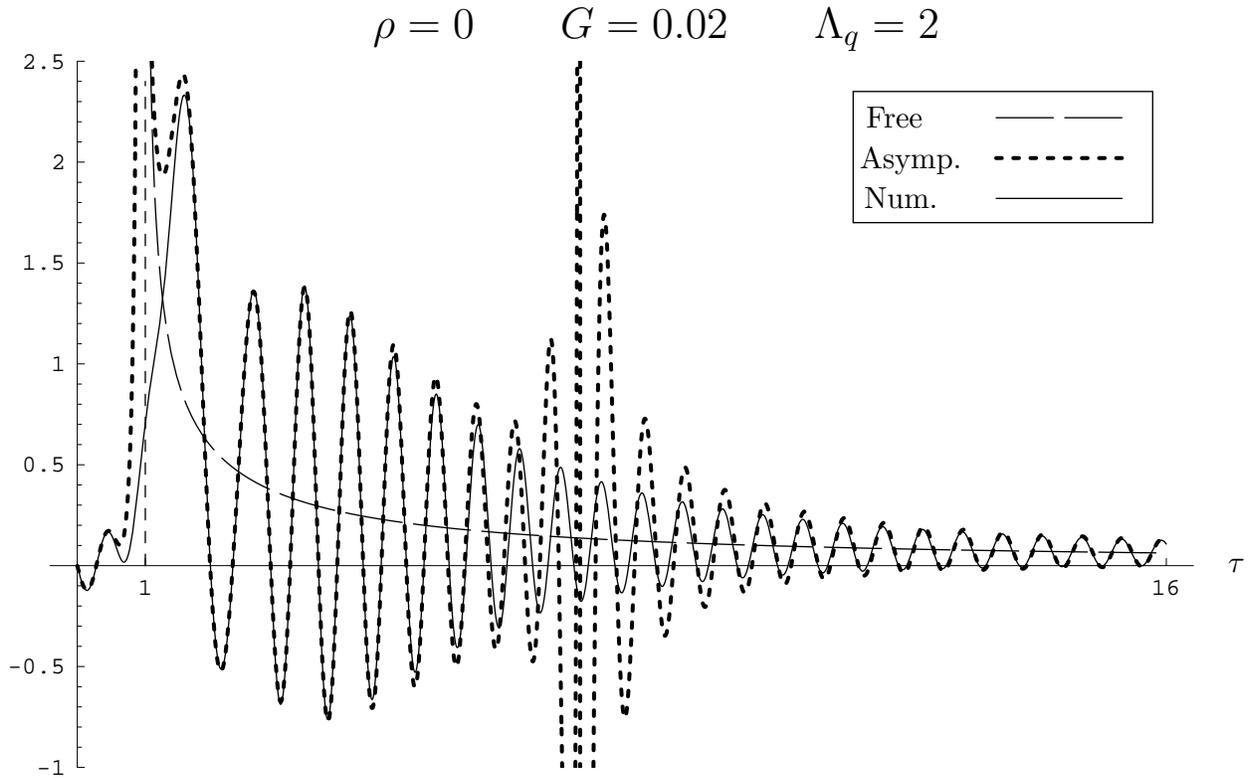}
\caption{Asymptotic approximation in $\lambda$ for the field
commutator over $8iG$ as a function of $\tau$ for $\rho=0$,
$G=0.02$, and $\Lambda_q=2$. We compare it {both} with a numerical
computation of the integral that defines it and {with} the
{unregulated} free commutator. As we can see{,} the
{approximation} is {good} except at the points where the
asymptotic behavior changes ($\tau=1$ {and} $\tau=e^{\Lambda_q}$).
Notice the difference in the amplitude of the oscillations for
$\tau<e^{\Lambda_q}$ and $\tau>e^{\Lambda_q}$.}
\label{fig1:propeje}
\end{figure}
\begin{figure}
\hspace{0cm}\includegraphics[width=16.5cm]{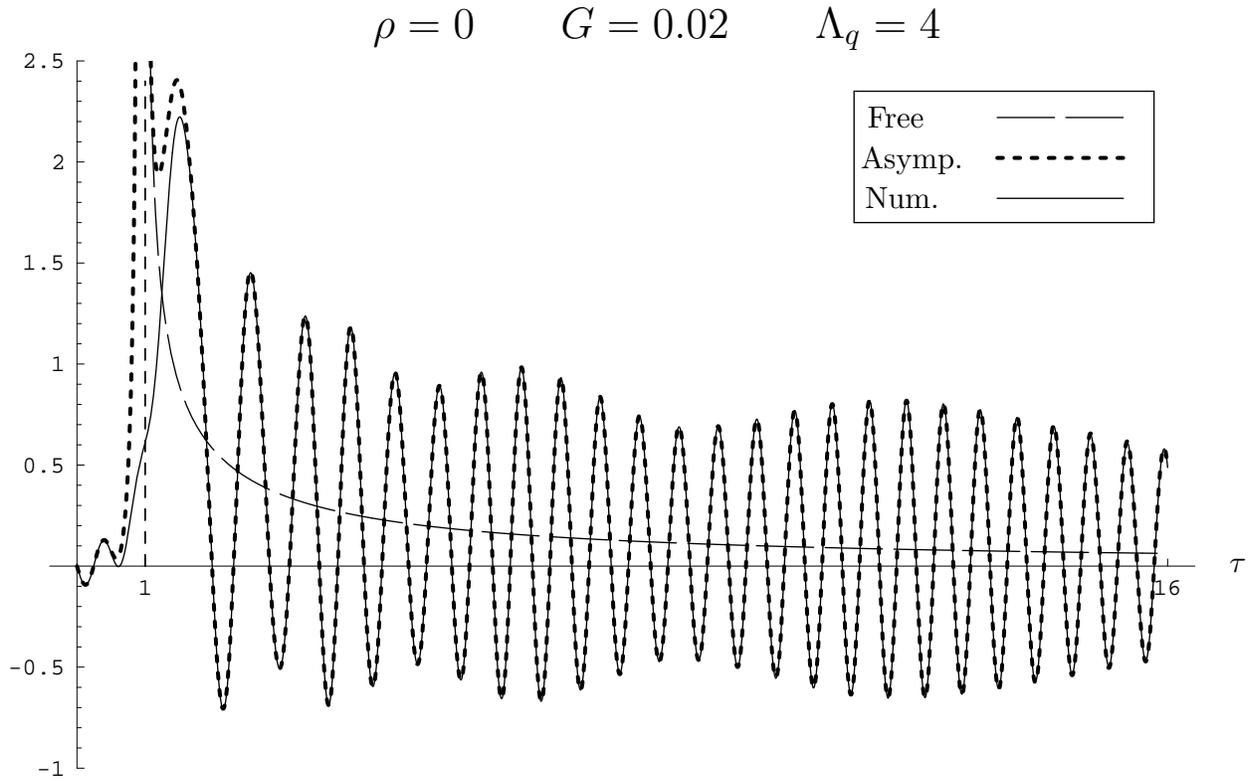}
\caption{Asymptotic approximation in $\lambda$ for the field
commutator over $8iG$ as a function of $\tau$ for $\rho=0$,
$G=0.02$, and $\Lambda_q=4$ (compared {both} with a numerical
computation of the integral that defines it and {with} the
{unregulated} free commutator). As we can see{,} the asymptotic
approximation is {good} except at $\tau=1$. Notice that the
cut-off introduces a modulation of the amplitude in the region
$1<\tau<e^{\Lambda_q}$. The singularity of the asymptotic
approximation at $\tau=e^{\Lambda_q}$ lies outside the plotted
region.}\label{fig1:propejeb}
\end{figure}
\begin{figure}
\hspace{0cm}\includegraphics[width=16.5cm]{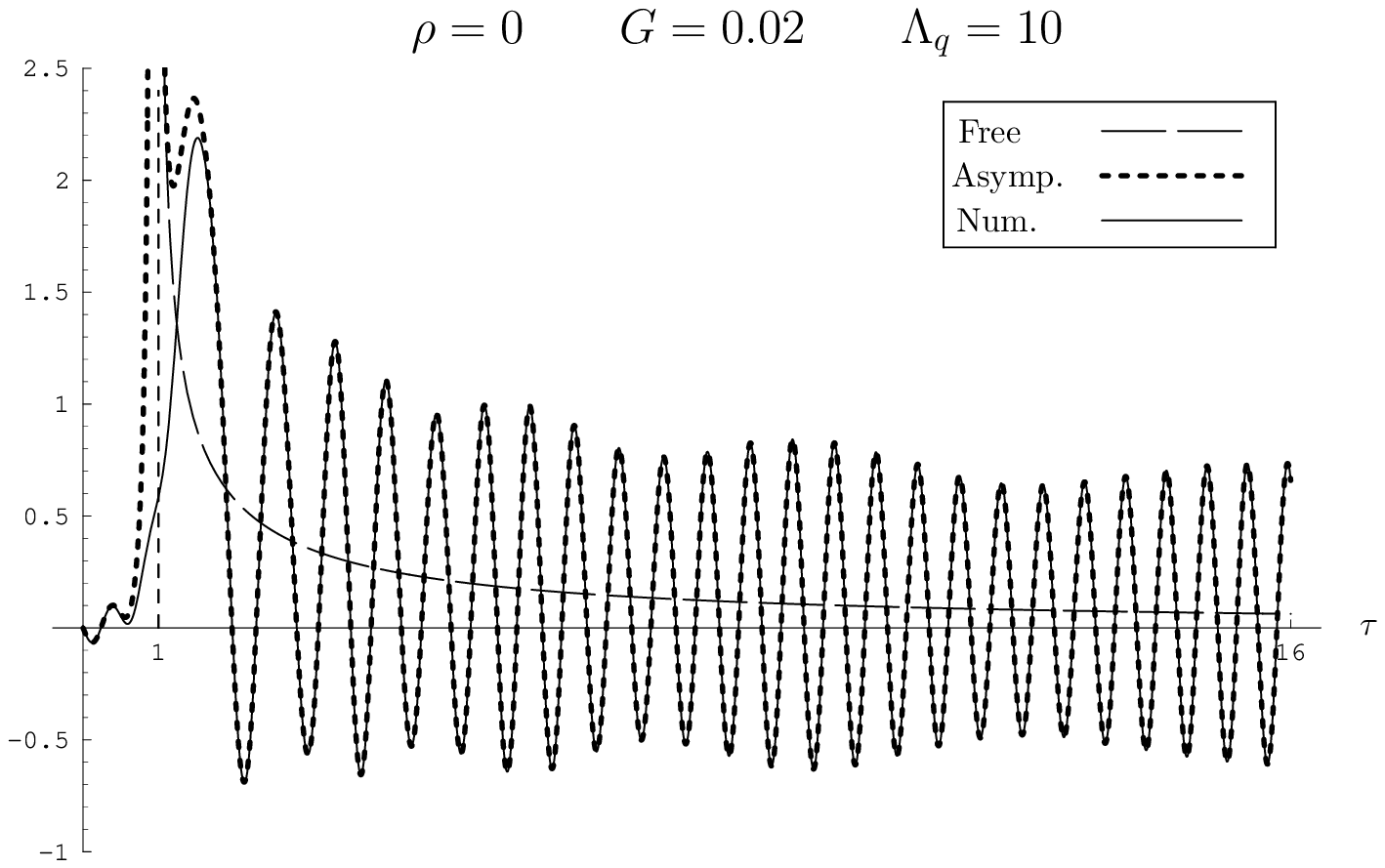}
\caption{Asymptotic approximation in $\lambda$ for the field
commutator over $8iG$ as a function of $\tau$ for $\rho=0$,
$G=0.02$, and $\Lambda_q=10$ (compared {both} with a numerical
computation of the integral that defines it and {with} the
{unregulated} free commutator). As we can see{,} the asymptotic
approximation is {good} except at $\tau=1$. The asymptotic
approximation obtained in Ref. \cite{BarberoG.:2004} is good in a
large region in the $\tau$ axis. The singularity of the asymptotic
approximation at $\tau=e^{\Lambda_q}$ lies outside the plotted
region.} \label{fig1:propejec}
\end{figure}
Figures \ref{fig1:propeje}-\ref{fig1:propejec} show the behavior
of the field commutator (over $8iG$) when $\rho=0$ as a function
of $\tau$ for several values of $\Lambda_q$. As we can see{,} the
asymptotic approximation becomes singular between regions with
different asymptotic regimes{,} but approximates {well} the exact
value of the commutator (obtained by numerical methods) for the
remaining values of $\tau$. Notice that the singularity at
$\tau=e^{\Lambda_q}$ of the asymptotic expansion lies outside the
plotted region in {Figs.} \ref{fig1:propejeb} and
\ref{fig1:propejec}.

In order to study the $\rho\neq0$ case we start by writing the
r.h.s. of (\ref{newintegral}) as
\begin{equation}
\Im\Big\{-\frac{\lambda
e^{i\tau\lambda}}{4\pi^2R_1}\int_0^{\Lambda_q}dq\oint_{\gamma_1}
dt_1\oint_{\gamma_2}dt_2\frac{1}{t_1t_2}
e^{\lambda\big[\frac{\scriptstyle
q}{\scriptstyle2}(t_1-\frac{\scriptstyle1}{\scriptstyle
t_1})+\frac{\scriptstyle \rho
q}{\scriptstyle2}(t_2-\frac{\scriptstyle1}{\scriptstyle
t_2})-i\tau e^{-q}\big]}{\Big\}}\label{trintegral}
\end{equation}
after {employing} the usual integral representation for the Bessel
functions $J_n$ ($n=0,\,1,\,\ldots$){.} Again it is {helpful} to
introduce the same neutralizers as above to split the integral in
three pieces $I_j$, $j=1,\,2,\,3$.

The integral $I_1$ gives the following two contributions{:}
\begin{equation}
\Im\bigg\{\frac{1}{2\pi^2R_1}
\oint_{\gamma_1}dt_1\oint_{\gamma_2}dt_2\frac{1}{\rho
t_1(t_2^2-1)+t_2(t_1^2+2i\tau t_1-1)}{\bigg\}}\label{freefront}
\end{equation}
and
\begin{equation}
\Im\bigg\{-\frac{2i\tau}{\pi^2R_1\lambda}
\oint_{\gamma_1}dt_1\oint_{\gamma_2}dt_2\frac{t_1^2t_2^2}{[\rho
t_1(t_2^2-1)+t_2(t_1^2+2i\tau t_1-1)]^3}\bigg\}.\label{doble}
\end{equation}
Both integrals can be computed exactly \cite{BarberoG.:2004} in
terms of complete elliptic integrals of first and second kind. The
first one gives the contribution of the {unregulated} free
commutator {(i.e., with} infinite cut-off{)}.
\begin{figure}
\includegraphics[width=8.5cm]{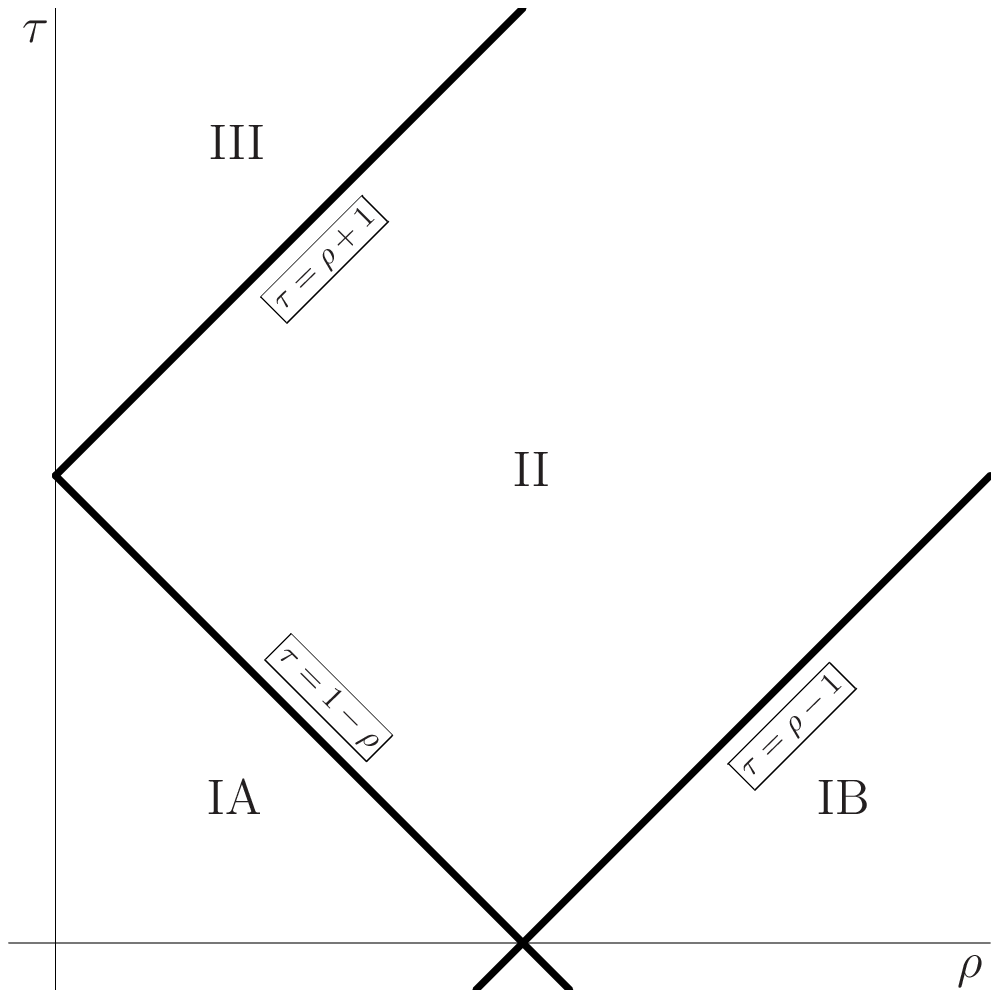} \caption{Regions in the
$(\rho,\tau)$ plane used in the discussion of the $\lambda$
asymptotics and the free commutator. Region I is defined by
$0<\tau<|\rho-1|$, region II by $|\rho-1|<\tau<\rho+1$, and region
III by $\rho+1<\tau$.} \label{fig4:regions}
\end{figure}
In order to describe it we define the regions I, II, and III by
$0<\tau<|\rho-1|$, $|\rho-1|<\tau<\rho+1$, {and} $\rho+1<\tau$
respectively. They are shown in {Fig.} \ref{fig4:regions}. In
region I the free commutator is zero, whereas in regions II and
III it is given by{:}

\bigskip

\noindent Region II

$$
\frac{1}{\pi
R_1\sqrt{\rho}}K\left(\sqrt{\frac{\tau^2-(\rho-1)^2}{4\rho}}\right),
$$

\noindent Region III

$$
\frac{2}{\pi R_1}\frac{1}{\sqrt{\tau^2-(1-\rho)^2}}
K\left(\sqrt{\frac{4\rho}{\tau^2-(1-\rho)^2}}\right).
$$

\noindent The second {contribution (\ref{doble})} can be computed
by the method outlined in {Appendix} IV of Ref.
\cite{BarberoG.:2004}{, obtaining:}

\bigskip

\noindent Region I

\begin{eqnarray}
&&\frac{\tau}{2\pi R_1\lambda}\left\{
\frac{2[1+\rho^4+2\tau^2-3\tau^4+2\rho^2(\tau^2-1)]\sqrt{(1+\rho)^2-\tau^2}}
{(1+\rho-\tau)^2(1-\rho+\tau)^2(-1+\rho+\tau)^2(1+\rho+\tau)^2}
E\left(\sqrt{\frac{4\rho}{(1+\rho)^2-\tau^2}}\right)\right.\nonumber\\
&&\hspace{2.0cm}\left.-\frac{2\tau^2}
{[\rho^4+(\tau^2-1)^2-2(1+\tau^2)\rho^2]\sqrt{(1+\rho)^2-\tau^2}}
K\left(\sqrt{\frac{4\rho}{(1+\rho)^2-\tau^2}}\right)\right\}
\end{eqnarray}

\noindent Region II

\begin{eqnarray}
&&\frac{\tau}{2\pi R_1\lambda}\left\{
\frac{1-2\rho^2+\rho^4+\tau^2-2\rho\tau^2+\rho^2
\tau^2-2\tau^4}{\sqrt{\rho}[(1-\rho)^2-\tau^2][(1+\rho)^2-\tau^2]^2}
K\left(\sqrt{\frac{(1+\rho)^2-\tau^2}{4\rho}}\right)\right.\nonumber\\
&&\hspace{1.1cm}\left.+\frac{4\sqrt{\rho}
[1-2\rho^2+\rho^4+2\tau^2-3\tau^4+2\rho^2\tau^2]}
{[\rho^4+(\tau^2-1)^2-2(1+\tau^2)\rho^2]^2}
E\left(\sqrt{\frac{(1+\rho)^2-\tau^2}{4\rho}}\right)\right\}{.}
\end{eqnarray}

\noindent The value of (\ref{doble}) in Region III is zero.

The integral $I_3${, on the other hand,} can {also} be studied by
the methods {described} in {Ref.} \cite{BarberoG.:2004}. The first
relevant term to its {asymptotic} expansion in inverse powers of
{$\lambda$} is derived from the double contour integral
\begin{equation}
\Im\bigg\{\frac{e^{i\tau\lambda}}{4\pi^2R_1}
\oint_{\gamma_1}dt_1\oint_{\gamma_2}dt_2\frac{2}{t_1t_2}
\frac{[t_1-\frac{1}{\displaystyle
t_1}+\rho(t_2-\frac{1}{\displaystyle t_2})+ 2i\tau
e^{-\Lambda_q}]e^{\lambda[\frac{{\Lambda_q}}{2}(t_1-\frac{1}{t_1})+
\rho\frac{{\Lambda_q}}{2}(t_2-\frac{1}{t_2})- i\tau
e^{-\Lambda_q}]}}{\Lambda_q^2[(t_1+\frac{1}{\displaystyle t_1})^2+
\rho^2(t_2+\frac{1}{\displaystyle
t_2})^2]-[t_1-\frac{1}{\displaystyle t_1}+
\rho(t_2-\frac{1}{\displaystyle t_2})+2i\tau
e^{-\Lambda_q}]^2}\bigg\}{,}\label{borderLambda}
\end{equation}
corresponding to $q=\Lambda_q$ {and} whose asymptotic behavior can
be {determined} by employing standard techniques for multiple
integrals \cite{Handels}. {In this} way we obtain the following
contribution{:}
\begin{eqnarray}
\frac{1}{2\pi R_1{\Lambda_q} \lambda
\sqrt{\rho}}\bigg\{\frac{\sin[\lambda\Lambda_q(1+\rho)-
\tau\lambda(1-e^{-\Lambda_q})]}{1+\rho-\tau
e^{-\Lambda_q}}-\frac{\sin[\lambda\Lambda_q(1+\rho)+
\tau\lambda(1-e^{-\Lambda_q})]}{1+\rho+\tau
e^{-\Lambda_q}}\hspace{5cm}&&\label{asimptbord}\\
\hspace{4cm}+\frac{\cos[\lambda\Lambda_q(1-\rho)-
\tau\lambda(1-e^{-\Lambda_q})]}{1-\rho-\tau
e^{-\Lambda_q}}-\frac{\cos[\lambda\Lambda_q(1-\rho)+
\tau\lambda(1-e^{-\Lambda_q})]}{1-\rho+\tau
e^{-\Lambda_q}}\bigg\}.&&\nonumber
\end{eqnarray}

\begin{figure}
\includegraphics[width=8.5cm]{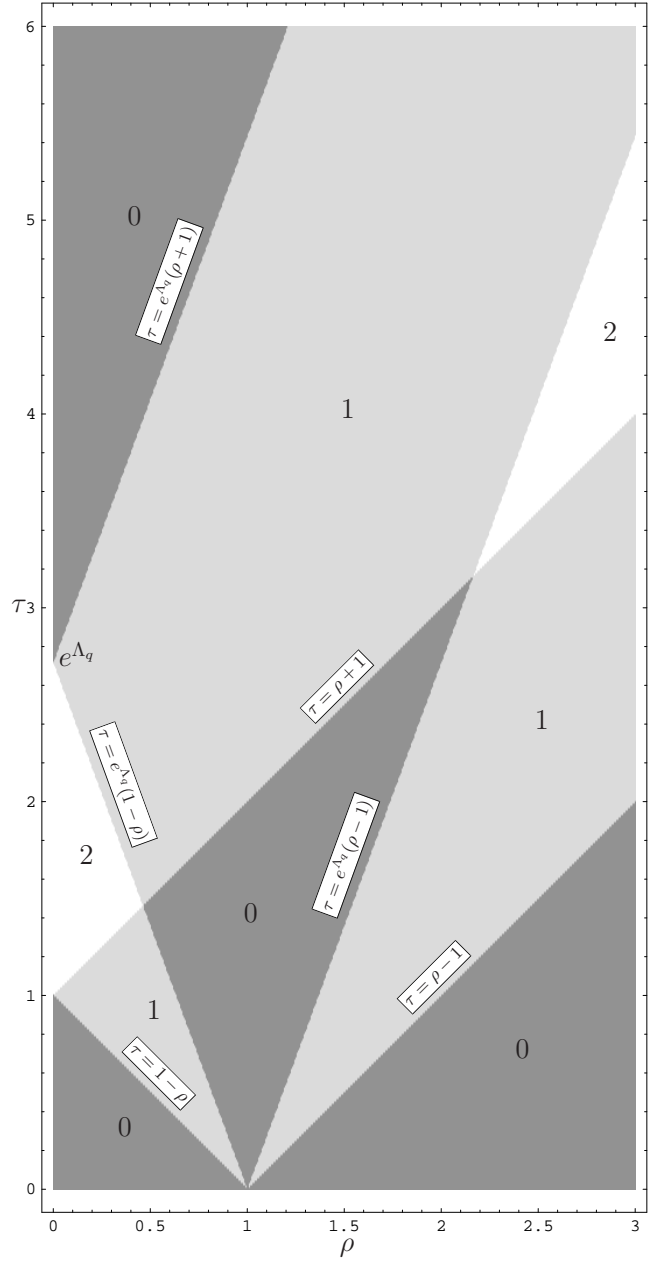} \caption{Regions in the
$(\rho,\tau)$ plane used in the discussion of the $\lambda$
asymptotics in the presence of a cut-off {$\Lambda_q=1$.} The
label of each region indicates how many critical points contribute
to the asymptotic expansion in $\lambda$.} \label{fig3:regcutoff}
\end{figure}

Finally{,} the integral $I_2$ is written in terms of a neutralizer
that {vanishes at} $q=0$ and $q=\Lambda_q$. This integral is best
studied by choosing the unit circumference centered in the origin
of the complex plane as the integration contour $\gamma$. The
contributions of this integral to the asymptotics of
(\ref{trintegral}) come from the stationary points of the exponent
in the integrand whenever they are within the integration region.
This {fact} is controlled by the value of the cut-off $\Lambda_q$.
The result is
$$
\{\theta(\tau-\rho+1)\theta[(\rho-1)e^{\Lambda_q}-\tau]\theta(\rho-1)+
\theta(\tau+\rho-1)\theta[(1-\rho)e^{\Lambda_q}-\tau]\theta(1-\rho)\}
\Im\bigg\{\frac{e^{-i\frac{\pi}{4}}
e^{i\lambda[\tau+|\rho-1|(1+\log\frac{\tau}{|\rho-1|})]}}
{R_1\sqrt{2\pi\lambda
\rho|1-\rho|}\log\frac{\tau}{|1-\rho|}}\bigg\}$$
$$
{+}\theta(\tau-\rho-1)\theta[(\rho+1)e^{\Lambda_q}-\tau]\Im\bigg\{
\frac{e^{i\frac{\pi}{4}}e^{i\lambda[\tau+(\rho+1)(\log\frac{1+\rho}{\tau}-1)]}}
{R_1\sqrt{2\pi\lambda
\rho(1+\rho)}\log\frac{\tau}{1+\rho}}\bigg\},\hspace{8cm}
$$
where the step functions define the regions where the different
stationary points contribute. As we can see { and it is explained
in {Fig.} \ref{fig3:regcutoff},} there are two contributions in
some parts of the $(\rho,\tau)$ plane, only one in some {other
parts,} and no contribution in the remaining ones.
Notice that{, whenever they differ from zero,} these contributions
are dominant with respect to {those} coming from the other
integrals.

Several points are now in order. First it is interesting to
realize that the singularity at ${\rho}=1$ that exists when the
cut-off is taken to be infinite (and is obviously {absent} now)
shows up as the region defined by the lines
$\tau=e^{\Lambda_q}(1-\rho)$, $\tau=e^{\Lambda_q}(\rho-1)$, {and}
$\tau=e^{\Lambda_q}(\rho+1)$ shrinks with growing $\Lambda_q$.
{Another interesting feature of the commutator when the cut-off is
present is the appearance of some regions where} {the leading
asymptotic behavior is not given by the expressions obtained in
Ref. \cite{BarberoG.:2004} for infinite cut-off, namely the
regions with $\tau>|1-\rho|$ labeled 0 in Fig.
\ref{fig3:regcutoff} and the region labeled 1 that connects them.
On the contrary,} {there are two regions where two stationary
points contribute to the asymptotics} {just} {as in the
$\Lambda_q\rightarrow\infty$ case}{, showing the characteristic
slow decay in the $\tau$ direction}\footnote{Notice that the other
contributions are subdominant with respect to this one{.}}{. Of
these two regions,} the one closer to the axis is bounded{,}
whereas the second one {[}defined by the lines
$\tau=e^{\Lambda_q}(\rho-1)$ and $\tau=\rho+1${]} is not. The
effect of the symmetry axis is evident in the sense that {it is
precisely there where} one of the lines that limits the boundary
of this region {starts, namely} $\tau=\rho+1$.

\begin{figure}
%\includegraphics[width=16.5cm]{figura4.eps}
%activar la linea anterior para mas resolucion y quitar la siguiente
\includegraphics[width=16.5cm]{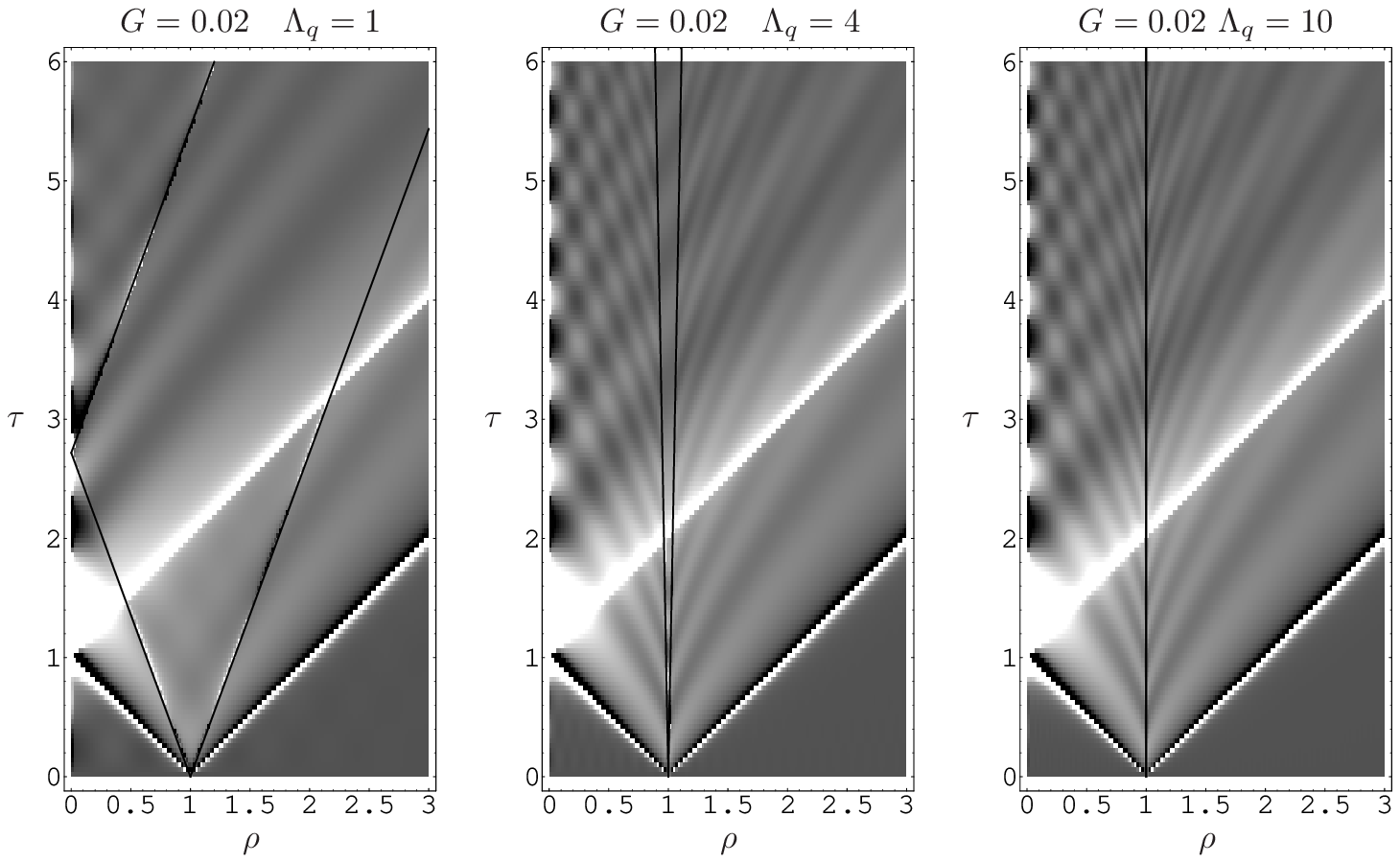}
\caption{Density plots of the commutator for $G=0.02$ and
different values of $\Lambda_q$. {Comparing the results with those
of Ref. \cite{BarberoG.:2004}, we} can see that the commutator in
the regions {labeled} 2 in {Fig.} \ref{fig4:regions} is
essentially equal to the one corresponding to {an} infinite
cut-off. Notice also the process by which the singularity at $R_1$
appears.} \label{fig5:regcutoffdens}
\end{figure}

\begin{figure}
\hspace{0cm}\includegraphics[width=16.5cm]{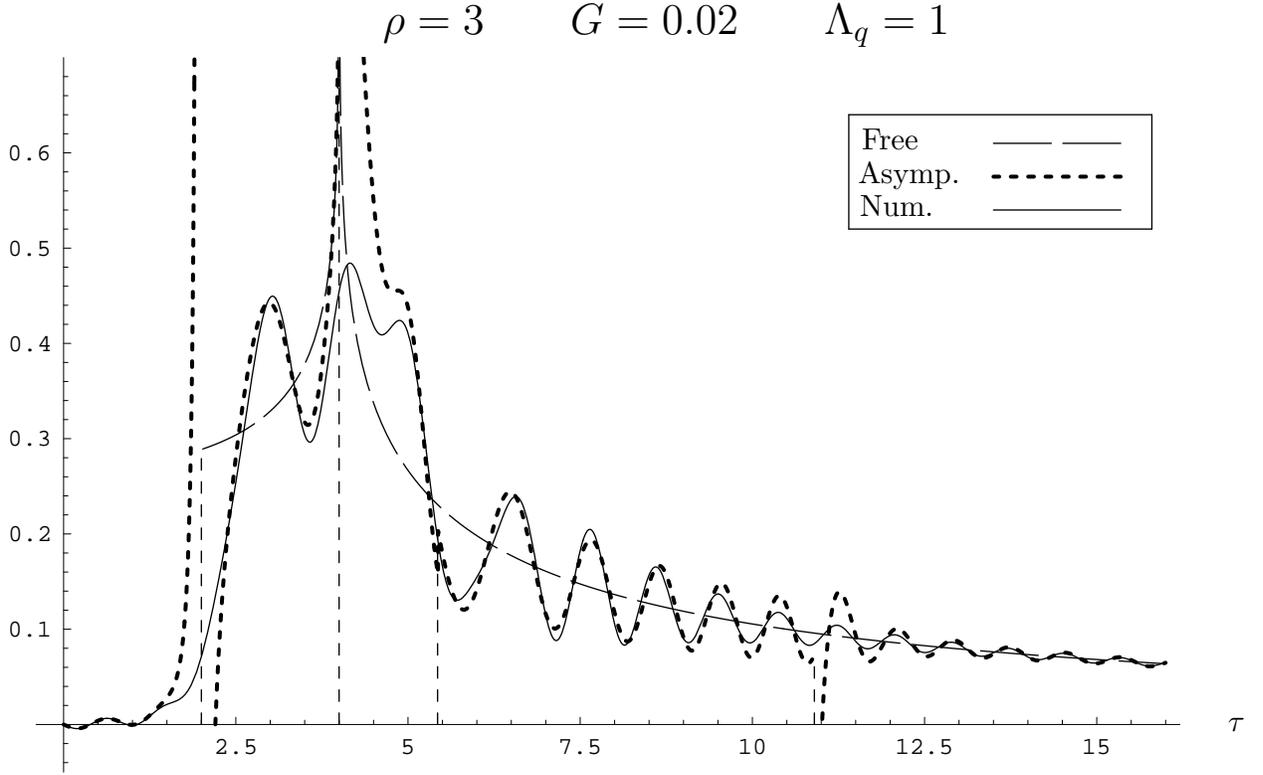}
\caption{Asymptotic approximation in $\lambda$ for the field
commutator over $8iG$ as a function of $\tau$ for $G=0.02$,
$\rho={3}${,} and $\Lambda_q=1$ compared with a numerical
approximation. The regions with different asymptotic regimes are
shown. {The} asymptotic approximation is {good} except at the
boundaries between these regions. The different types of
{behavior} are also {evident}.} \label{fig6:propneq0-1}
\end{figure}

\begin{figure}
\hspace{0cm}\includegraphics[width=16.5cm]{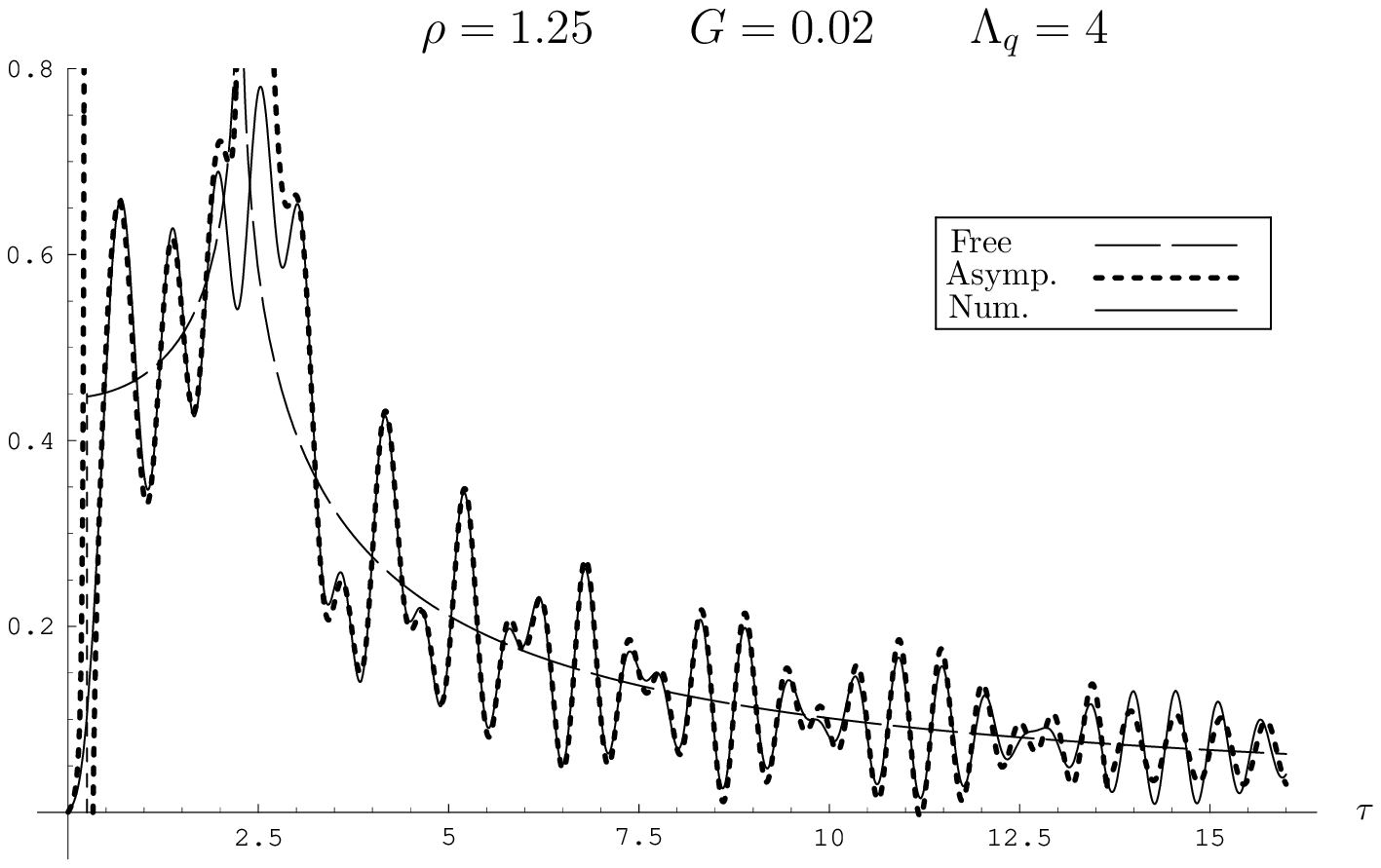}
\caption{Asymptotic approximation in $\lambda$ for the field
commutator over $8iG$ as a function of $\tau$ for $G=0.02$,
$\rho=1.25${,} and $\Lambda_q=4$ compared with a numerical
approximation. } \label{fig6:propneq0-2}
\end{figure}

\begin{figure}
\hspace{0cm}\includegraphics[width=16.5cm]{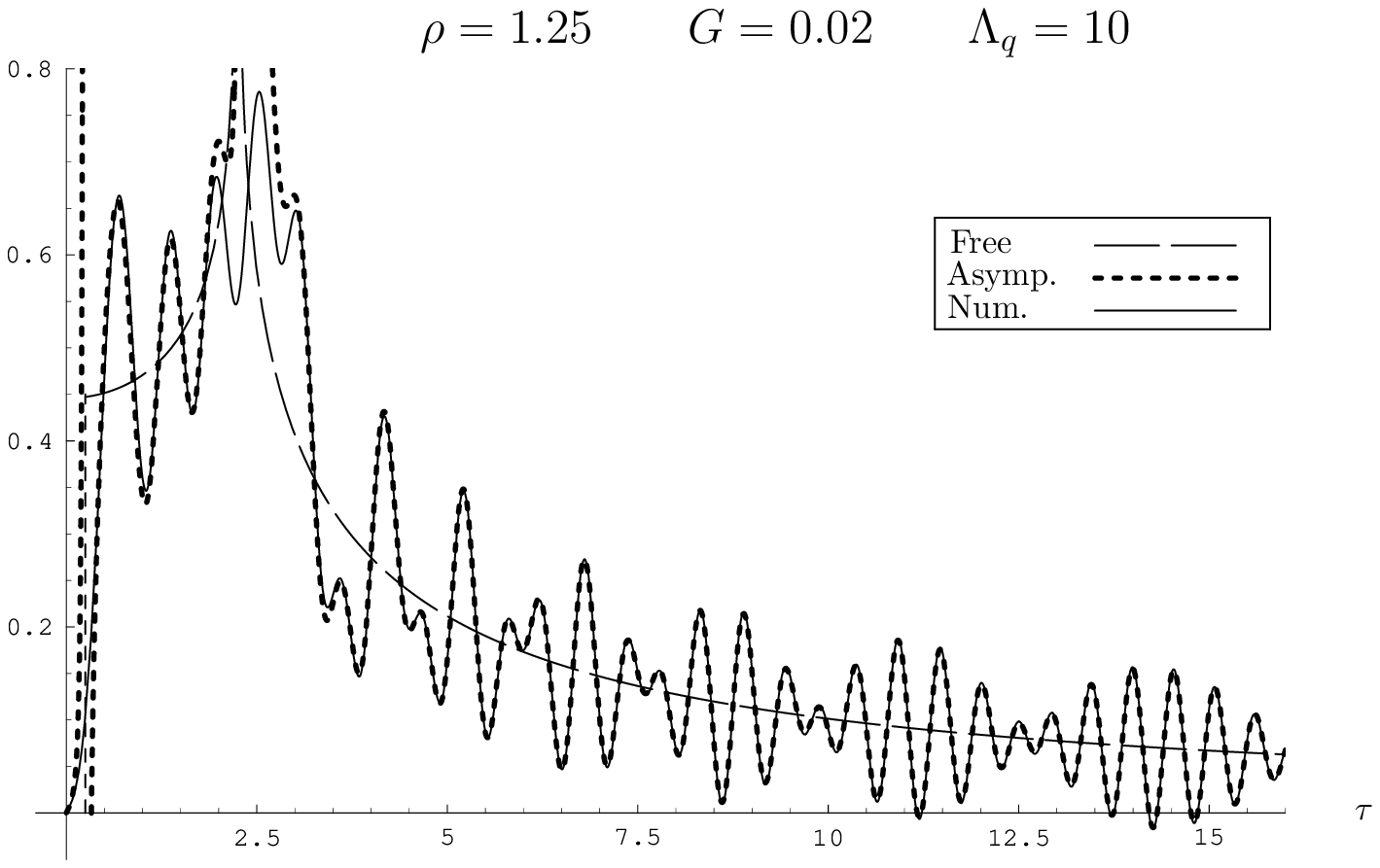}
\caption{Asymptotic approximation in $\lambda$ for the field
commutator over $8iG$ as a function of $\tau$ for $G=0.02$,
$\rho=1.25${,} and $\Lambda_q=10$ compared with a numerical
approximation.} \label{fig6:propneq0-3}
\end{figure}

As we can see{, the influence} of the cut-off is important in some
parts of the ($\rho,\tau$) {plane, }but there are others where the
asymptotic behavior is described {at leading order(s)} by the
{unregulated} $\lambda\rightarrow\infty$ limit. The {consideration
of these different} regions {helps in describing} the intermediate
regimes where the infinite cut-off approximation is expected to
work, at least for {large} values of $\lambda$. Finally we want to
point out that the most dramatic {quantum} effect {observed} when
the cut-off is infinite --the very slow {fall-off} of the
commutator at the axis in the $\tau$ direction-- is no longer
present {after introducing a regulator}. This casts some {doubts}
about the ``observability" of large quantum gravitational
{fluctuations} at the axis. These behaviors can be {visually}
appreciated in {Figs.}
\ref{fig5:regcutoffdens}-\ref{fig6:propneq0-3}{.}


{
\section{\label{power}Power Expansion in $G$}
}

In the {above} sections we have discussed the asymptotics of the
regulated {field} commutator as a function of $\rho$, $\lambda$,
and $\tau$. These are dimensionless parameters obtained by using
$R_1$ as a length scale. {We want to discuss now the possibility
of expanding} this commutator as a power series in $G$. The main
{motivation} to consider this {issue is that} one would expect to
{arrive at an} expansion {of this kind when adopting a standard
perturbative approach for the} treatment of the problem. As we
will see{,} this can be done in a rather straightforward way if a
cut-off is introduced {in the system. However,} our {description
breaks down} when the cut-off is removed.

Let us analyze then the expansion of the vacuum expectation value
of the commutator in powers of the quantum gravitational
constant\footnote{In previous papers
\cite{BarberoG.:2003ye,BarberoG.:2004} we chose to expand
$\sin[kt\frac{(1\!-\!e^{-q})}{q}]$ as a power series in $e^{-q}$.
This yielded a convergent series representation for the
commutator. Nonetheless, the complicated dependence of this
expansion in $G$ did not allow us to obtain a power series in {the
gravitational constant}.} $G=G_3\hbar$. To this end we rewrite
(\ref{integral}) as
\begin{equation}
\frac{1}{8iG}\langle 0|\,
[\hat{\psi}_E(R_1,t_1),\hat{\psi}_E(R_2,t_2)]\,|0\rangle=
\int_0^{\Lambda_k}\!\!\!\!\!\!dk\,J_0(R_1 k)J_0(R_2
k)\sin\left[k{R_1\tau }\frac{(1\!-\!e^{-q})}{q}\right]\!.
\hspace*{.4cm}\label{vacom}
\end{equation}
With our conventions, both $k R_1\tau$ and ${q=Gk}$ are
dimensionless, whereas $k$ {can be regarded to have} dimensions of
an inverse length. Note that all the dependence on $G$ is
contained in $q${, accepting that the cut-off $\Lambda_k$ is
fixed}. Thus, in order to arrive at the desired series, we will
expand the integrand in powers of {the} variable {$q$. At this
point, it is worth remarking that, had we described the regulated
commutator by means of the dimensionless cut-off
$\Lambda_q=4G\Lambda_k$ as in previous sections\footnote{{Let us
comment that one can always make dimensionless the cut-off
$\Lambda_k$ (as well as $k$ and $1/G$) by multiplication with an
arbitrary, fixed length scale independent of the gravitational
constant, so that no spurious dependence on $G$ is introduced in
the process. In this sense, one may always view $\Lambda_k$ either
as a $G$-independent dimensionful cut-off or equivalently as a
fixed dimensionless parameter.}}, it would not have been possible
to single out the dependence on the gravitational constant via
that on $q$.}

We will use the following formulas for the Taylor expansion of the
functions involved in our expression {(\ref{vacom})} and the
composition of the resulting series, assuming for the moment their
convergence:
\begin{eqnarray}\label{expa}
g(q)&:=&\frac{1-q-e^{-q}}{q}=\sum_{n=1}^{\infty}\frac{(-q)^n}{(n+1)!},\\
\label{sina} \sin{\left(kR_1\tau+y\right)}&=&\sin{\left(k
R_1\tau\right)}\sum_{m=0}^{\infty}(-1)^m
\frac{y^{2m}}{(2m)!}+\cos{\left(kR_1\tau\right)}\sum_{m=0}^{\infty}(-1)^m
\frac{y^{2m+1}}{(2m+1)!},\\ \label{compo}
{[}g(q){]}^m&=&\left[\sum_{n=1}^{\infty}
\frac{(-q)^n}{(n+1)!}\right]^m=\sum_{p=m}^{\infty}a_p[m]\,(-q)^p,\\
\label{apms}
a_p[m]&:=&\sum_{\sigma(p|m)}\prod_{i=1}^{m}\frac{1}{(n_i+1)\,!}\,,\end{eqnarray}
where the range of the last sum extends to the sets of
$m$-integers $n_i$ given by \begin{equation}\label{sigma}
\sigma(p|m):=\left\{n_i\geq 1;\, \sum_{i=1}^{m}
n_i=p\right\}.\end{equation} Interchanging the sum and integration
orders, one then obtains the formal series
\begin{eqnarray}\label{vaser}
\frac{1}{8iG}\langle
0|[\hat{\psi}_E(R_1,t_1),\hat{\psi}_E(R_2,t_2)]|0 \rangle=\!
\sum_{p=0}^{\infty}\int_0^{{\Lambda_k}}\!\!dkJ_0(R_1k)J_0(R_2k)(-4Gk)^p
\,F_p(kR_1\tau),\end{eqnarray} with
\begin{eqnarray}\label{fms}
\!\!\!\!&&F_p(kR_1\tau):=\sin{(kR_1\tau)}\sum_{m=1}^{{\rm
int}[p/2]}\frac{(-1)^m}{(2m)!}a_p[2m]\,(kR_1\tau)^{2m}\nonumber\\
&&\hspace{2cm}{+}\cos{(kR_1\tau)}\sum_{m=0}^{{\rm
int}[(p-1)/2]}\frac{(-1)^m}{(2m+1)!}a_p[2m+1]\,(kR_1\tau)^{2m+1},\;\;\;
\;p\geq 1,\\ \!\!\!\!&&F_0(kR_1\tau):=\sin{(kR_1\tau)}\nonumber.
\end{eqnarray}
Here, the function ${\rm int}[x]$ is the integer part of $x$, and
the sum over $m$ that multiplies the function $\sin{(kR_1\tau)}$
is understood to vanish when $p=1$. Note that{, in the case of
infinite cut-off,} the first ($p=0$) term reproduces the
commutator of the free-field theory. {Moreover, then all} the
$p\geq1$ additions to the free field contribution are integrals
over $[0,\infty)$ of oscillating, non-bounded functions and,
hence, at best conditionally convergent. {So, in the unregulated
theory ($\Lambda_k=\infty$),} the above expansion should be taken
only as a formal expression, and therefore we expect that the
{corresponding} vacuum expectation value of the field commutator
is not analytic in $G$.

Of course these problems disappear {when we admit the existence of
a finite cut-off $0<\Lambda_k<\infty$.} Taking into account that
all the functions $F_p(kR_1\tau)$ are analytic in $k$ around the
positive real axis, because $F_p$ is a finite combination of
products of analytic functions, and that so are the zeroth-order
Bessel functions that appear in the integrals of (\ref{vaser}), it
is easy to conclude that all those integrals are well-defined when
they are restricted to a compact interval $[0,\Lambda_k]$. {Thus,
each term in the power series (\ref{vaser}) is finite for any
finite positive value of $\Lambda_k$.}

In the rest of this section, we will discuss the formal
manipulations that we have carried out with infinite sums in order
to deduce the above expansion. First, notice that the Taylor
series in {(\ref{expa})}, which is obtained from that of the
exponential function, has an infinite convergence radius. When
this series is substituted in (\ref{vacom}), one obtains a
trigonometric function similar to that on the l.h.s. of
{(\ref{sina})}, but with $y=kR_1\tau \,g(q)$ expanded in powers of
$q$. On the other hand, relation (\ref{sina}) is just the formula
for the sine of the sum of two angles, with the resulting
functions $\sin{y}$ and $\cos{y}$ {replaced with} their Taylor
expansion. The series compositions $\sin{[kR_1\tau g(q)]}$ and
$\cos{[kR_1\tau g(q)]}$ can then be rearranged without problems
employing for ${[}g(q){]}^m$ the value given in {(\ref{compo})}
because $g(q)$ (that we recall that converges for all
$q\in\mathbb{R}^+$) is always smaller than the convergence radii
of the sine and cosine series, which are in fact infinite.

In this way, one arrives at a expectation value of the {regulated}
commutator that is equal to an integral over the interval $k\in
[0,\Lambda_k]$ of the series of functions $\sum_{p}
f_p(k|R_1,R_2,{\tau})$, with
\begin{equation}\label{fmrs}
f_p(k|R_1,R_2,{\tau}):=J_0(R_1k)J_0(R_2k)(-4Gk)^p F_p(k{R_1}\tau).
\end{equation}
Since the functions $f_p(k)$ are clearly continuous in
$k\in[0,\Lambda_k]$ (for all allowed values of $R_1$, $R_2$ and
$\tau$) and this interval is compact, they are all integrable in
that region. As a consequence, it is sufficient that the
considered series of functions converges uniformly in
$k\in[0,\Lambda_k]$ to guarantee that the integration can be
interchanged with the infinite sum. We will postpone to the next
section the proof of this uniform convergence, at least for a
convenient choice of the cut-off.

In conclusion, we have seen that the {field} commutator in vacuo,
regulated with a {(dimensionful)} fixed cut-off, can be expanded
as a power series in the gravitational constant $G$, each term in
the series being finite. Besides, all the manipulations performed
to deduce this series are rigorously justified provided that the
cut-off is chosen so that the series $\sum_{p}
f_p(k|R_1,R_2,{\tau})$ converges uniformly in $k\in[0,\Lambda_k]$.
Furthermore, in fact this requirement of uniform convergence
automatically ensures that the corresponding integrated power
series (\ref{vaser}) converges, and that it does so to the actual
value of the expectation value of the regulated
commutator.\newline

\section{Uniform convergence}

We want to demonstrate that there exists a non-zero value of the
cut-off for which the series $\sum_{p} f_p(k|R_1,R_2,{\tau})$
converges uniformly in $k\in[0,\Lambda_k]$ for any fixed
{non-negative value of $R_1$, $R_2$, and $\tau$}. Let us start by
finding a convenient upper bound for the coefficients $a_p[m]$,
with $m\geq 1$, defined in {(\ref{apms})}. First, note that
$a_p[m]=0$ unless $p\geq m$, because no set of the form
$\sigma(p|m)$ {exists} with $n_i\geq 1$ if $\sum_{i=1}^{m}n_i=p<
m$. In addition, since $(n_i+1)\,!\geq 2$ for $n_i\geq 1$, we have
that
\begin{equation}\label{apmb} a_p[m]\leq
\frac{1}{2^m}\,\sum_{\sigma(p|m)}1.\end{equation}

From our definition (\ref{sigma}), the last sum equals the
different ways to arrange $p-m$ non-distinguishable elements
[namely, the excess about its minimum of the sum {of} $m$ elements
$n_i\geq 1${, which equals $p-m$} for $\sigma(p|m)$] between $m$
different sets [which correspond to the $m$ integers $n_i$]. The
result is given by the permutations of $(p-m)+m-1$ elements [the
latter $m-1$ elements representing movable delimiters between the
$m$ sets] with possible repetition in $p-m$ (the genuine,
non-distinguishable elements) and in $m-1$ (the imaginary
delimiters). Thus,
\begin{equation}\label{apmb2} a_p[m]\leq
\frac{1}{2^m}\,\frac{(p-1)\,!}{(p-m)\,!(m-1)\,!}\leq
\frac{1}{2^m}\,\frac{p\,!}{(p-m)!}.\end{equation} In the last
inequality we have employed that $(m-1)!\,p\geq 1$ for all $p\geq
m\geq 1$.

Using that the absolute value of the sine and the cosine is never
greater than the unity, it is not difficult then to deduce from
{(\ref{fms})} the following bound for $F_p(k{R_1}\tau)$, with
$p\geq 1$,
\begin{equation}\label{fmsb}
\left|F_p(k{R_1}\tau)\right|\leq \sum_{m=1}^p
\frac{k^m{R_1\tau}^m}{m\,!}\,a_p[m]\leq \sum_{m=1}^p
\left(\frac{k{R_1\tau}}{2}\right)^m
\left(\begin{array}{c}p\\m\end{array}\right)=\left(1+\frac{k{R_1
\tau}}{2}\right)^p -1.\end{equation} In the last step, we have
employed the formula of the binomial expansion. Likewise, since
the zeroth-order Bessel function is bounded by the unity in the
positive real axis, we get that, for all {non-negative} values of
$R_1$ and $R_2$,
\begin{equation}\label{fmrsb}
\left|f_p(k|R_1,R_2,{\tau})\right|\leq(4Gk)^p
\left[\left(1+\frac{k{R_1\tau}}{2}\right) ^p-1\right]\leq
\left[4Gk\left(1+\frac{k{R_1\tau}}{2}\right)
\right]^p.\end{equation} The last inequality is trivial, given
that $4Gk\geq 0$. Note also that the bound on the r.h.s. is valid
even in the case $p=0$, taking into account {(\ref{fmrs})}.

Finally, since $4Gk(1+k{R_1\tau}/2)$ is a strictly increasing
function of $k$ in $[0,\Lambda_k]$, we obtain a bound independent
of the variable $k$ in the interval considered,
\begin{equation}\label{fmrsb2}
\left|f_p(k|R_1,R_2,{\tau})\right|\leq
\left[4G\Lambda_k\left(1+\frac{\Lambda_k{R_1\tau}}{2}\right)
\right]^p.\end{equation} To obtain the desired convergence
properties, it will suffice to require that
\begin{equation}\label{pert}
4G\Lambda_k\left(1+\frac{\Lambda_k{R_1\tau}}{2}\right)< 1.
\end{equation} We then get that both
$\left|f_p(k|R_1,R_2,{\tau})\right|$ and its integral over
$[0,\Lambda_k]$ are small corrections for large $p$, which tend to
zero in the limit $p\rightarrow\infty$.

In order to prove the uniform convergence of the series
$\sum_{p}f_p(k|R_1,R_2,{\tau})$, we have to check that, for each
$\epsilon>0$, there exists an integer $P$ such that, for every
$k\in[0,\Lambda_k]$,
\begin{equation}\label{unico}
\left|\sum_{p=P}^{\infty}f_p(k|R_1,R_2,{\tau})\right|\leq
\epsilon.
\end{equation}
Taking into account inequality (\ref{pert}), it is clear that,
given $\epsilon>0$, we can always find a sufficiently large
integer $P$ for which
\begin{equation}\label{pertp}
\left[4G\Lambda_k\left(1+\frac{\Lambda_k{R_1\tau}}
{2}\right)\right]^P<\epsilon\left[1-4G\Lambda_k
\left(1+\frac{\Lambda_k{R_1\tau}}{2}\right) \right].
\end{equation}
Note that the choice of this $P$ depends only on the values of
$\epsilon$, $\Lambda_k$, $G$ and ${R_1\tau}$. Using
{(\ref{fmrsb2})}, we then have
\begin{eqnarray}\label{unib}
\left|\sum_{p=P}^{\infty}f_p(k|R_1,R_2,{\tau})\right|\leq
\sum_{p=P}^{\infty}\left[4G\Lambda_k\left(1+\frac{\Lambda_k
{R_1\tau}}{2}\right) \right]^p =
\frac{\left[4G\Lambda_k\left(1+\frac{\Lambda_k{R_1\tau}}{2}
\right)
\right]^P}{1-4G\Lambda_k\left(1+\frac{\Lambda_k{R_1\tau}}{2}
\right)}\leq \epsilon.\nonumber
\end{eqnarray}
So, inequality (\ref{unico}) is valid for all $k$ in the
considered interval $[0,\Lambda_k]$, as we wanted to prove.

We have thus shown that, for a given $\tau$, every choice of the
cut-off $\Lambda_k>0$ that satisfies condition (\ref{pert}) leads
to a convergent power series in the gravitational constant $G$ for
the expectation vacuum of the {regulated} commutator, regardless
of the radial coordinates $R_1$ and $R_2$. Moreover, the power
expansion converges indeed to the true value of this {regulated}
commutator in vacuo.

\section{Conclusions and comments}

We have studied in this paper {the issue of} microcausality for
quantum Einstein-Rosen waves after a suitable cut-off is
introduced to {regulate} the quantum fields. {In more detail, we
have considered the introduction of a momentum cut-off $\Lambda_k$
(or its dimensionless counterpart $\Lambda_q$).} We have
{discussed} first the asymptotic expansions in terms of the
dimensionless parameters $\rho$, $\tau$, and $\lambda$ along the
lines of {Ref.} \cite{BarberoG.:2004}. Owing to the fact that
{these} parameters are defined with the help of $R_1${, in
principle one does not need to make explicit} the dependence of
the cut-off {$\Lambda_k$ on $G$} in this case{. On physical
grounds, one could view this cut-off}, for example, as the inverse
of the Planck length.

{We} have seen {that} the introduction of a {finite} cut-off
{modifies} some of the conclusions obtained in {Ref.}
\cite{BarberoG.:2004}. In particular we have seen that some of the
most dramatic effects present {when the cut-off is infinite} (in
particular the behavior of the field commutators in the symmetry
axis) are {now} somewhat mitigated{.} Nevertheless we have been
able to show that the approximation provided by the {unregulated
field commutator} is a good one in some regions {of} the
$(\rho,\tau)$ plane, and, in fact, there is an unbounded region
where that approximation prevails. This indicates that{, even
though the influence} of the cut-off is felt in some regions of
the parameter space{, it is} irrelevant {in} others.

In {Secs. VI and VII, on the other hand,} we have considered {the
expansion of the field commutator} in {terms of} the gravitational
constant $G$. We notice, {nonetheless,} that condition
(\ref{pert}) on the cut-off {$\Lambda_k$}, that guarantees the
convergence of the series, depends on $G$. {At this stage, one}
possibility would be to admit that the cut-off depends on the
gravitational constant; however, the expansion obtained would
{then} fail to provide a genuine power series in $G$, because this
parameter would {also} enter the different terms in the series via
the implicit dependence of $\Lambda_k$ on it. {Another}
possibility that {indeed} respects the interpretation of our
expansion as a power series in $G$ is the following. Employing
that {condition (\ref{pert}) is} an inequality equation for
$\Lambda_k$ {given} in terms of a function of $G$ that is strictly
increasing, it is easy to see that the inequality is satisfied for
all values of $G$ in a certain interval $[0,G_M]$ if and only if
it is satisfied for $G_M$. Something similar happens with respect
to the dependence on the value of ${R_1}\tau={|t_2-t_1|}$, so that
if we want to consider a whole time interval of the form
${|t_2-t_1|\in[0,t_M]}$ , we only have to evaluate our condition
at the maximum time lapse. In other words, to ensure the
convergence of the series for $G\in[0,G_M]$ and any time
difference in $[0,t_M]$, we only have to demand the requirement
(\ref{pert}) at $G=G_M$ and ${R_1\tau}=t_M$, because then
\begin{equation}\label{pertm}
4G\Lambda_k\left(1+\frac{\Lambda_k{R_1\tau}}{2}\right)<
4G_M\Lambda_k\left(1+\frac{\Lambda_k{t_M}}{2}\right)< 1.
\end{equation}

In this way we arrive at a cut-off that is independent of the
particular values considered for ${|t_2-t_1|}$ and the
gravitational constant (in the commented intervals), and our
expansion becomes a true power series in $G$. The above inequality
leads to the following positive upper bound for $\Lambda_k$:
\begin{equation}\label{cutgt}
\Lambda_k\leq\frac{1}{{t_M}}\left(\sqrt{1+\frac{{t_M}}
{2G_M}}-1\right).
\end{equation}
Therefore, with a cut-off that satisfies this condition, the power
series (\ref{vaser}) converges in the interval $[0,G_M]$ for all
radial positions $R_1$ and $R_2$ and
${R_1\tau=|t_2-t_1|\in[0,t_M]}$.

When ${t_M}$ is small, the bound on $\Lambda_k$ is approximately
$1/(4G_M)$, whereas for large ${t_M}$ it is nearly equal to
$1/\sqrt{2G_M{t_M}}$. In particular, with this bound the cut-off
would have to be vanishingly small if we want a good convergent
behavior in an infinitely large time interval
(${t_M}\rightarrow\infty$). An open question is whether it is
possible or not to find a different, non-zero time-independent
cut-off such that the expansion of the {regulated} commutator
converges for any value of the time elapsed, i.e. for {all}
${|t_2-t_1|} \in\mathbb{R}^+$. We expect to encounter convergence
problems when the time interval is unbounded; for instance, one
can prove that the series (\ref{vaser}) does not converge
uniformly in ${\tau}\in\mathbb{R}^+$ with any choice of the
cut-off $\Lambda_k$ (for generic $R_1$ and $R_2$). Nonetheless,
one can in fact consider a kind of semi-classical limit in which
$G_M$ tends to zero (and hence so does the value of the
gravitational constant, that had been restricted to $[0,G_M]$),
while the time interval where the convergence is granted reaches
infinity.

In order to do this, one only needs to allow a dependence of
${t_M}$ on $G_M$, so that the assumed maximum value of the time
difference varies with that of the gravitational constant.
Suppose, let us say, that ${t_M}(G_M)=G_M^{\,-\alpha}$ with
$0<\alpha<1$. Then, the bound (\ref{cutgt}) on the cut-off becomes
\begin{equation}\label{cutgg}
\Lambda_k\leq
G_M^{\,\alpha}\left(\sqrt{1+\frac{1}{2G_M^{\,(\alpha+1)}}}
-1\right).
\end{equation}
Thus, when $G_M$ tends to zero, we get the asymptotic behavior
$\Lambda_k\leq G_M^{\,(\alpha-1)/2}/\sqrt{2}$. Since
$G_M^{\,(\alpha-1)/2}$ and $G_M^{\,-\alpha}$ diverge for vanishing
$G_M$, because $0<\alpha<1$, we therefore conclude that the
cut-off can be removed in the limit $G_M\rightarrow 0$ while
ensuring that the time interval $\left[0,{t_M}(G_M)\right]$, where
the expansion is well-defined, covers the positive real axis.

We finally discuss the physical interpretation of this type of
cut-off. It turns out to be intimately related to the maximum
resolution that can be reached for the physical time when a
{certain} perturbative approach is adopted to describe the quantum
dynamics \cite{BarberoG.:2003at}. In such an approach, one expands
the evolution generator in powers of $G$ and regards the
free-field Hamiltonian as the dominant contribution, with the
higher powers seen as corrections. The auxiliary time $T$,
associated with the free-field Hamiltonian, plays then the role of
evolution parameter in the quantum theory, whereas the physical
time becomes an operator. It was shown in Ref.
\cite{BarberoG.:2003at} that, under these circumstances, a
resolution {limit} $\Delta t$ emerges for the physical time,
\begin{equation}\label{treso}
[\Delta t]^2\geq 4G^2+4G T.\end{equation}

Employing the inequality $\sqrt{1+x}\leq x/(\sqrt{1+x}-1)$ for
$x>0$, evaluated at ${x=t_M/(2G)}$, one can easily check from
condition (\ref{cutgt}) that the inverse of the cut-off satisfies
\begin{equation}\label{cireso}
\Lambda_k^{-1}\geq\sqrt{4G_M^{\,2}+2G_M t_M}.\end{equation}
Therefore, {the bound on} $\Lambda_k^{-1}$ equals {that on} the
time resolution $\Delta t$ for a value $G=G_M$ of the
gravitational constant and a time elapsed $T=2t_M$ (and thus of
the same order as $t_M$). {In} this sense, one can assign to
$\Lambda_k^{-1}$ the interpretation of a genuine resolution limit
in the physical time.

The future prospects for this line of work will {focus} on the
issue of deriving and obtaining meaningful physical information
from the {$S$} matrix of the model. We feel that the mathematical
techniques employed here to study the asymptotics of field
commutators, with and {without} a cut-off{,} will be {also helpful
in analyzing this issue}. We plan to concentrate on this problem
in the future.

\begin{acknowledgments}
This work was supported by the Spanish {MEC} under the research
projects BFM2001-0213 and BFM2002-04031-C02-02. {The authors want
to thank A. Ashtekar, J. A. Torresano, and M. Varadarajan for
helpful comments, as well as J. Klauder for an interesting
question.}
\end{acknowledgments}

%\bibliography{lambda}

\end{document}